\documentclass[12pt]{article}
\usepackage{amsmath,amssymb,bbold,hyperref,graphicx,color}
\newcommand{\be}{\begin{equation}}
\newcommand{\ee}{\end{equation}}
\newcommand{\bea}{\begin{eqnarray}}
\newcommand{\eea}{\end{eqnarray}}
\newcommand{\beas}{\begin{eqnarray*}}
\newcommand{\eeas}{\end{eqnarray*}}

\begin{document}
\begin{titlepage}

\begin{center}

{\Large Bulk reconstruction using timelike entanglement in (A)dS}

\vspace{12mm}

\renewcommand\thefootnote{\mbox{$\fnsymbol{footnote}$}}
Avijit Das${}^{1}$\footnote{adas73439@gmail.com},
Shivrat Sachdeva${}^{2}$\footnote{shivratsachdeva@hri.res.in} and
Debajyoti Sarkar${}^{3}$\footnote{dsarkar@iiti.ac.in}

\vspace{6mm}

${}^1${\small \sl Department of Physics} \\
{\small \sl Central University of Karnataka} \\
{\small \sl Kadaganchi, Karnataka 585367, India}

\vspace{2mm}
${}^2${\small \sl Harish-Chandra Research Institute} \\
{\small \sl Chhatnag Road, Jhunsi, Allahabad 211019, India}

\vspace{2mm}
${}^3${\small \sl Department of Physics} \\
{\small \sl Indian Institute of Technology Indore} \\
{\small \sl Khandwa Road 453552 Indore, India}

\end{center}

\vspace{12mm}

\noindent
It is well-known that the entanglement entropies for spacelike subregions, and the associated modular Hamiltonians play a crucial role in the bulk reconstruction program within Anti de-Sitter (AdS) holography. Explicit examples of HKLL map exist mostly for the cases where the emergent bulk region is the so-called entanglement wedge of the given boundary subregion. However, motivated from the complex pseudo-entropy in Euclidean conformal field theories (CFT), one can talk about a `timelike entanglement' in Lorentzian CFTs dual to AdS spacetimes. One can then utilize this boundary timelike entanglement to define a boundary `timelike modular Hamiltonian'. We use constraints involving these Hamiltonians in a manner similar to how it was used for spacelike cases, and write down bulk operators in regions which are not probed by an RT surface corresponding to a single CFT. In the context of two dimensional CFT, we re-derive the HKLL formulas for free bulk scalar fields in three examples: in AdS Poincar\'{e} patch, inside and outside of the AdS black hole, and for de Sitter flat slicings. In this method, one no longer requires the knowledge of bulk dynamics for sub-horizon holography.

\end{titlepage}
\setcounter{footnote}{0}
\renewcommand\thefootnote{\mbox{\arabic{footnote}}}

\hrule
\tableofcontents
\bigskip
\hrule

\addtolength{\parskip}{8pt}

\section{Introduction\label{sect:intro}}

Quantum entanglement finds many applications in condensed matter and high energy theories alike, and it is an inherent property of many quantum systems including discrete many body physics and also continuum field theory. Our work in the present paper will discuss a very particular application of entanglement, namely its usefulness in understanding gravitational systems in the context of gauge/ gravity duality. More concretely, given a \emph{holographic} conformal field theory (CFT), how the entanglement structure of the said CFT leads to an \emph{emergent} bulk geometry with one extra dimension, which is often a geometry containing asymptotically Anti de Sitter (AdS) spacetime.   

The present topic has been studied rigorously during the last decade culminating in a thorough understanding of how spacelike entanglement leads to bulk Einstein dynamics at the linearized and higher orders. By spacelike entanglement we mean that given the knowledge of a field theory state on a spacelike Cauchy slice $\Sigma$, we can find the corresponding reduced density matrix $\rho_A$ for a given subregion $A$ of the spacelike surface (in our case, $\Sigma=A\,\cup\, A^c$, where $A^c$ is the complement region). This density matrix can in turn compute the statistical von-Neumann entropy which is usually taken as the measure of the entanglement of the above subregion with the rest.\footnote{This statement is already marred with a rich plethora of subtleties involving partitionability of field theory Hilbert space, due to its continuum nature and for any gauge symmetries that might be present. We will ignore all these subtleties for this study. For a recent review of these aspects, see \cite{Witten:2018lha}. For example, for a field theory, a better defined quantity is the extended modular Hamiltonian $\tilde{H}$, which, if the system were bi-partitionable like above, would give
\begin{equation}\label{e	q:extmodhamsl}
	\tilde{H}_A=H_A-H_{A^c}\,.
\end{equation}
} The entanglement entropy is given by 
\begin{equation}
	S_{EE}=-\text{Tr}(\rho_A\log\rho_A)=\text{Tr}(\rho_A H_A)\,,
\end{equation}
where $H_A$ is the so-called subregion modular Hamiltonian or entanglement Hamiltonian. As is clear from the equation above, it is an operator-valued measure of entanglement entropy, thus carrying all the details of the entanglement structure of the mixed state in question. 

We would be interested in applying these rich ideas in the bulk reconstruction program of AdS/CFT duality. Once again, the history of bulk reconstruction is quite long, almost spanning the lifetime of the duality itself, but for concreteness, we will be confining ourselves with the Lorentzian extrapolate boundary-to-bulk dictionary. A typical example of such map was provided by Hamilton et al. \cite{Hamilton:2005ju,Hamilton:2006az,Hamilton:2006fh}, following whom it is currently known as the HKLL map. In its original form, the HKLL map provides the bulk to boundary kernel (by solving bulk field equations and imposing required boundary valued constraints) for numerous fields in various asymptotically AdS backgrounds \cite{Kabat:2012hp,Sarkar:2014dma,Roy:2015pga,Bhattacharjee:2022ehq}, which enables one to write a \emph{local} bulk field in terms of the dual boundary operators. To illustrate, for a given bulk field $\Phi(x)$, located at the bulk point $x$, one can write down a formula involving corresponding boundary operator $\mathcal{O}(X)$ supported over the intersection between the causal wedge of the bulk point $x$, and the boundary. 
\begin{equation}
	\Phi(x)=\int_{boundary} dX'\,K(X'|x)\,\mathcal{O}(X')+\mathcal{O}(1/N)+\mathcal{O}(\alpha')\,.
\end{equation}
The free bulk fields require evaluations of bulk dynamics in the original HKLL approach, whereas the corrections in CFT's central charge $N$, or string length $\alpha'$ can be incorporated by considering bulk microcausality.  As an interesting extension, using the bulk dynamics in de Sitter (dS), a similar map was written down in \cite{Xiao:2014uea} (this will also be a topic of our interest here, although the current status of a possible dS/CFT duality is much murkier than its AdS counterpart. See also \cite{Sarkar:2014jia}). For us, any possible dS/CFT subtleties will not be relevant, as we will simply assume that such a duality exists.

Later on, the HKLL formula was re-derived for scalar and gauge fields in \cite{Kabat:2017mun,Kabat:2018smf} from a purely algebraic perspective (for other relevant works along this theme, see \cite{Faulkner:2017vdd,Faulkner:2018faa,Roy:2018ehv}). Following the proposals of JLMS \cite{Jafferis:2015del}, which equates the bulk and boundary modular Hamiltonians within AdS/CFT duality, they use the predicted behavior of bulk fields under the flow of extended modular Hamiltonian. In particular, for a holographic CFT, the free bulk scalar fields are invariant under the modular flow, if they are located on the Ryu-Takayanagi (RT) surface \cite{Ryu:2006bv} corresponding to the subregion. Such a condition then naturally deliver the HKLL map as a byproduct. While this clearly demonstrates the power of spacelike entanglement (Hamiltonian) in answering questions regarding bulk gravity, there are a few loopholes that remain to be filled. The usual spacelike $\tilde{H}$ can only produce a modular flow within the entanglement wedge of the dual boundary subregion. Therefore, in the thermofield double states e.g., where the AdS black hole is an entangled product of two copies of maximally entangled CFTs \cite{Maldacena:2001kr}, any given boundary observer can at most probe up to the horizon of the black hole by means of a flow produced by their $\tilde{H}$. Indeed, so far, no explicit sub-horizon bulk reconstruction exists, which avoids using bulk equation of motion. We intend to fill this gap in the literature utilizing the timelike entanglement, and its ability to see the bulk regions way past the reach of $\tilde{H}$.

\subsection{Outline of the paper}

The plan of the paper is as follows. We start in section \ref{sect:intro2tlEE} by reviewing how to construct timelike entanglement entropy for two dimensional CFTs by starting from the spacelike entanglement entropies and performing some trivial Wick rotations. These Wick rotation techniques can also be applied at the level of replica trick to obtain the corresponding `timelike modular Hamiltonians'. This is the topic covered in section \ref{sect:intro2tlEH}. In both these two sections, we have covered the vacuum states at both zero and finite temperature. Later on, in appendix \ref{app:app} we have studied the flow under these boundary timelike modular Hamiltonians by mapping the Virasoro symmetry to the bulk. This above study already provides us with a solid hindsight for the boundary constraints that we use in the main part of the paper.

Next we turn to the bulk locality program using these timelike modular Hamiltonians (with a JLMS type relation working also in the timelike case). Sections \ref{sect:poincareoutside}, \ref{sect:btzout} and \ref{sect:inside} are devoted to bulk reconstruction in respectively Poincar\'{e} AdS vacuum state, outside the BTZ black hole \cite{Banados:1992wn}, and inside the BTZ black hole. Later, section \ref{sect:dS} deals with the bulk locality in dS flat slices. These results provide the main outcome of our work and illustrate the usefulness of the timelike modular Hamiltonian (and timelike entanglement in general), in the extrapolate dictionary. Finally, after a brief conclusion and discussions on the possible future avenues, we have a couple of extra appendices providing some technical results skipped during the main texts.

\section{Timelike entanglement entropy}\label{sect:intro2tlEE}

\subsection{CFT methods}

In this section we will provide a brief review of the timelike entanglement entropy (EE) that has been a topic of recent interest especially in the context of pseudo entanglement entropy and its variants. Some references that we found particularly useful are \cite{Nakata:2020luh,Doi:2022iyj,Narayan:2022afv,Li:2022tsv,Doi:2023zaf,Narayan:2023zen} (some earlier works discussing timelike entanglement include \cite{Wang:2018jva}, which were expanded further in \cite{Jiang:2023ffu,Jiang:2023loq}). There are various ways to derive the entanglement entropy for a timelike subregion within CFT, and in what follows we will briefly outline these various methods by closely following \cite{Doi:2022iyj,Doi:2023zaf}. This section contains published ideas, and has been included here to make the paper self-contained. The readers who are already familiar with these works, can skip ahead to section \ref{sect:intro2tlEH}.\footnote{It's natural to ask whether the timelike EE also follows from a density matrix or not. Indeed it does, and this density matrix is closely related to the transition density matrix discussed in the context of pseudo-entropy. Interested readers may want to look at \cite{Nakata:2020luh,Doi:2022iyj} for a brief discussion on this point.}

\subsubsection{Rotating the interval}

\begin{figure}[h]
    \centering
    \includegraphics[scale=0.5]{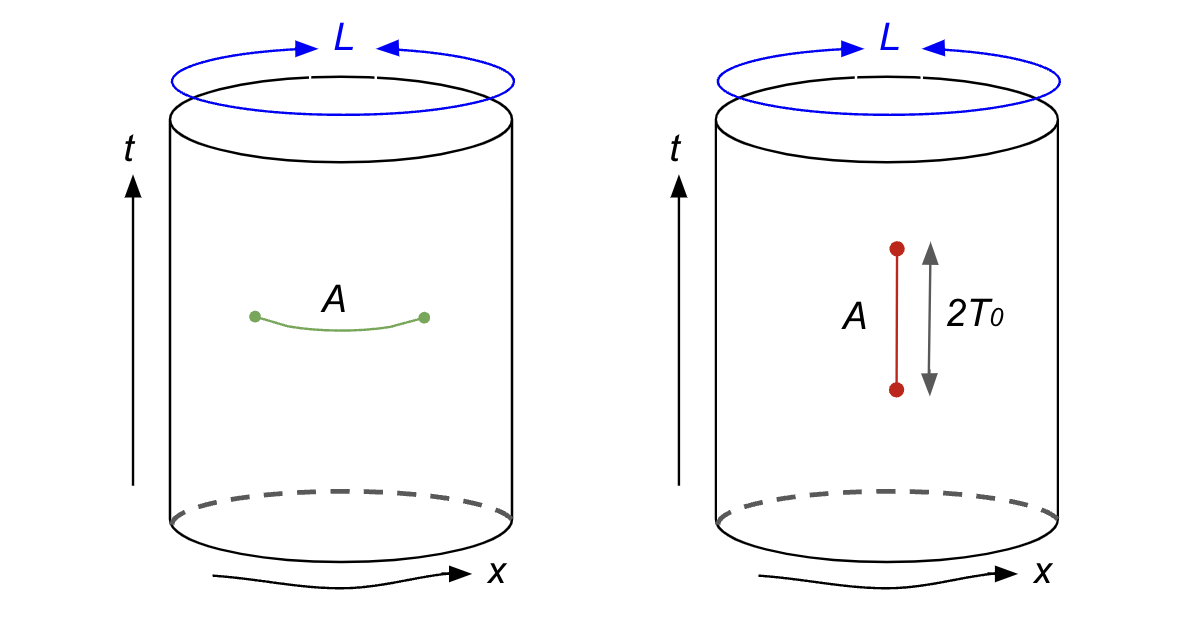}
    \caption{The green spatial region $A$ (left) has been rotated to make it a timelike red region (right). Under this rotation, the spacelike entanglement entropy goes over to its timelike counterpart.}
    \label{fig:TEE_01}
\end{figure}

As we review below, the simplest way to extract the timelike entanglement entropy is to start with the entanglement entropy for a general spacelike interval, and then performing the necessary spacetime rotations to make the interval timelike (for an illustration, applicable for the case (ii) below, see figure \ref{fig:TEE_01}). For example, one can start with the usual replica method to compute the  entanglement entropy (see e.g.~\cite{Cardy:2016fqc} for a review) and study a few well-known cases.

\vspace{1mm}

\noindent
\emph{(i) Intervals in a CFT$_2$ on a plane:} Let's begin with a spacelike subsystem $A$ with the two endpoints separated by $\Delta x$ and $\Delta t$ in spacetime. The $n$-th Renyi entropy $S_{A}^{(n)}$, is computed using the two-point correlation function of twist operators and in the limit \(n \to 1\), we arrive at the von Neumann entropy
\begin{equation}
    S_A = S_A^{(1)} = \frac{c}{3}\log\left(\frac{\sqrt{(\Delta x)^2 - (\Delta t)^2}}{\epsilon}\right).
\label{B}    
\end{equation}
Here $c$ represents the central charge of the CFT, with $\epsilon$ being the UV cutoff of the theory.

At this stage, to find out the timelike entanglement entropy $S_A^{(T)}$,\footnote{We will always use a superscript $(T)$ to denote the timelike quantities, and the non-superscripted ones will denote the usual spacelike counterparts.} in the case when the subsyetm $A$ is timelike, we can simply derive the resulting equation by making the interval $(\Delta x)^2 - (\Delta t)^2$ negative in equation (\ref{B}). Doing so, it provides a complex valued quantity given by 
\begin{equation}
   S_A^{(T)} = \frac{c}{3}\log\left(\frac{\sqrt{(\Delta t)^2 - (\Delta x)^2}}{\epsilon}\right) + \frac{c\pi i}{6}\,.
\label{C}    
\end{equation}

For most of our considerations below, we will be interested in a subsystem that is purely timelike with length $2T_0$ (i.e. with $\Delta x = 0$ and $\Delta t= 2T_0$). In that case, we have
\begin{equation}
   S_A^{(T)} = \frac{c}{3}\log\left(\frac{2T_0}{\epsilon}\right) + \frac{c\pi i}{6}\,.
\label{D}    
\end{equation}
\vspace{1mm}

\noindent
\emph{(ii) CFT$_2$ intervals on a periodic space:} On the other hand, if the CFT has a compact spatial direction (denoted by $\phi$ in this case) with circumference $L$ (at zero temperature), then the EE for such a  subsystem $A$ is
\begin{equation}
    S_A = \frac{c}{6}\log\left(\frac{L^2}{\pi^2\epsilon^2}\sin\left(\frac{\pi(\Delta\phi + \Delta t)}{L}\right)\sin\left(\frac{\pi(\Delta\phi - \Delta t)}{L}\right)\right)\,.
\label{E}    
\end{equation}
Once again, we can find out the timelike entanglement entropy $S_A^{(T)}$ by making the subsystem \(A\) timelike, i.e.~by imposing $(\Delta\phi - \Delta t) < 0$. We then obtain 
\begin{equation}
    S_A^{(T)} = \frac{c}{6}\log\left(\frac{L^2}{\pi^2\epsilon^2}\sin\left(\frac{\pi(\Delta t + \Delta\phi)}{L}\right)\sin\left(\frac{\pi(\Delta t - \Delta\phi)}{L}\right)\right) + \frac{i\pi c}{6}\,.
\label{F}    
\end{equation}
Again, for a purely timelike subsystem $A$ (i.e~with $\Delta\phi = 0$ and $\Delta t = 2T_0$) we have
\begin{equation}
    S_A^{(T)} = \frac{c}{3}\log\left(\frac{L}{\pi \epsilon}\sin\frac{2\pi T_0}{L}\right) + \frac{i\pi c}{6}\,.
 \label{G}   
\end{equation}
Of course, the above equation \eqref{G} boils down to \eqref{D} in the infinite $L$ limit.

\vspace{1mm}

\noindent
\emph{(iii) CFT$_2$ intervals at a finite temperature:} If the CFT is at a finite temperature $1/\beta$ on an infinite line, then for a spacelike subregion $A$, the corresponding entanglement entropy is well known ($t$ here is the Euclidean time) \cite{Wong:2013gua}: 
\begin{equation}
    S_A = \frac{c}{6}\log\left(\frac{\beta^2}{\pi^2\epsilon^2}\sinh\left(\frac{\pi(\Delta x + \Delta t)}{\beta}\right)\sinh\left(\frac{\pi(\Delta x - \Delta t)}{\beta}\right)\right)\,.
\label{H}    
\end{equation}
Once again, rotating the subregion so that it becomes timelike, we obtain the corresponding timelike entanglement entropy $ S_A^{(T)}$ as
\begin{equation}
    S_A^{(T)}= \frac{c}{6}\log\left(\frac{\beta^2}{\pi^2\epsilon^2}\sinh\left(\frac{\pi(\Delta t + \Delta x)}{\beta}\right)\sinh\left(\frac{\pi(\Delta t - \Delta x)}{\beta}\right)\right) + \frac{i\pi c}{6}\,.
\label{I}    
\end{equation}
Once more, for a purely timelike subsystem of length $2T_0$ we get
\begin{equation}
     S_A^{(T)} = \frac{c}{3}\log\left(\frac{\beta}{\pi\epsilon}\sinh\left(\frac{2\pi T_0}{\beta}\right)\right) + \frac{i\pi c}{6}\,.
\label{J}    
\end{equation}
The similarity of expressions between \eqref{J} and \eqref{G} is natural, and quite correctly, under the zero temperature limit $\beta\to\infty$, \eqref{J} reproduces \eqref{D}.

\subsubsection{Wick rotating the coordinates}\label{sec:wickee}

An alternative way of deriving the above equations for timelike entanglement entropies is via Wick rotating the coordinates \cite{Doi:2022iyj,Doi:2023zaf}. 
To be concrete, we would discuss here the case of a timelike interval for a CFT at zero temperature with a periodic spatial direction. We will discuss two different types of Wick rotations, as both of them appear in the literature and both of them clarify the similarity between the cases (ii) and (iii) above. Essentially, the Wick rotation turns the spatial coordinate timelike (and vice-versa), and hence the spacelike interval will now have the interpretation of a timelike interval.
\vspace{1mm}

\begin{figure}[h]
    \centering
    \includegraphics[scale=0.5]{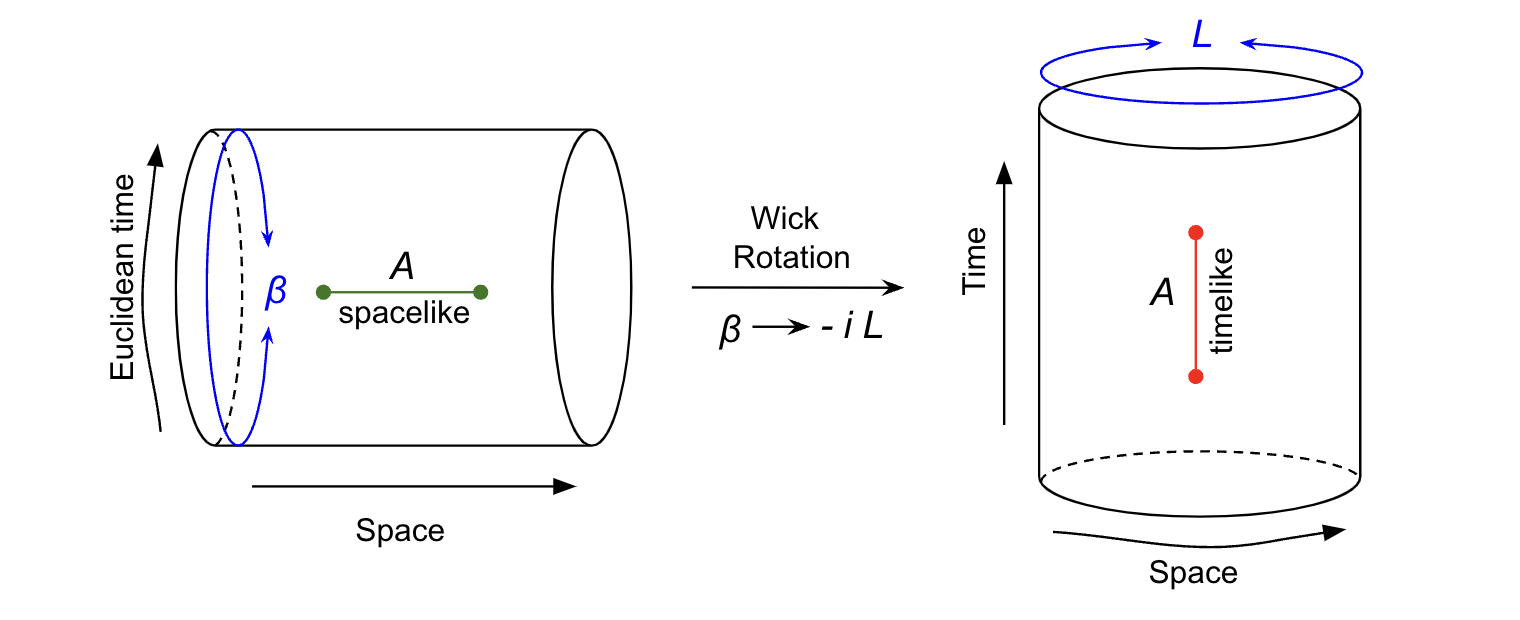}
    \caption{Timelike entanglement entropy via a Wick rotation.}
    \label{fig:wick_02}
\end{figure}

\noindent
\emph{(i) Wick rotation type I:} In this case, we can start with a CFT at a finite temperature $1/\beta$ on an infinite line. From \eqref{H} we know that for a spacelike subregion on a constant time slice, the resulting EE is given by 
\begin{equation}
    S_A = \frac{c}{3} \log \left( \frac{\beta}{\pi\tilde{\epsilon}} \sinh \left( \frac{2\pi R}{\beta} \right) \right)\,.
\label{V}    
\end{equation}
We have denoted here the size of the spacelike region to be $2R$, and the UV cut-off to be $\tilde{\epsilon}$. If we now perform the Wick rotation on the Euclidean time $\beta\to -iL$, it rotates the CFT cylinder so the resulting geometry has the interpretation of a zero temperature CFT on a periodically defined Lorentzian space. See figure \ref{fig:wick_02} for an illustration of this point.
However, we have to simultaneously perform a Wick rotation of the UV cut-off scale $\tilde{\epsilon}\to -i\epsilon$, so that $\epsilon$ is now the usual UV cut-off in the CFT. Under these analytic continuations, the subregion automatically becomes a timelike subregion, and we can identify the spatial interval $2R$ as the timelike interval $2T_0$. The resulting entanglement is precisely \eqref{G}:
\begin{equation}
    S_A^{(T)} = \frac{c}{3} \log \left( \frac{L}{\pi\epsilon} \sin \left( \frac{2\pi T_0}{L} \right) \right) + \frac{i\pi c}{6}\,.
\label{Z}    
\end{equation}

\vspace{1mm}

\noindent
\emph{(ii) Wick rotation type II:} Alternatively, we can start with the same system as in the left panel of figure \ref{fig:wick_02} and perform the Wick rotations as given below
\begin{equation}
    {x}\rightarrow{it} \qquad\text{and}\qquad R\to iT_0 \label{2}\,.
\end{equation}
Clearly the main difference here is that we have analytically continued the spatial coordinate only (so it now has an interpretation of timelike direction), but we had to supplement it with an explicit analytic continuation of the subregion (without having to complexify the UV cut-off $\epsilon$, and $\beta$ identified with the spatial periodicity $L$). Once again, under these changes, \eqref{V} boils down to \eqref{Z}.

Of course, we can also obtain the entanglement entropy for a timelike interval when the CFT is at a finite temperature. In that case, we will simply start with a spacelike interval for a CFT on a spatially periodic direction, and perform Wick rotations as discussed above.

\subsection{Holographic methods}

We will conclude our review of timelike EE by briefly discussing the associated bulk extremal surfaces if the CFT in question is holographic. These extremal surfaces will be important for us later when performing the bulk reconstruction. We will consider two examples, one being the pure AdS$_3$ and the other being the BTZ black hole.\footnote{We will not discuss the AdS$_3$-Rindler spacetimes separately from BTZ, as the analysis there essentially follows the BTZ analysis with minor, tractable differences.}

\subsubsection{Pure AdS$_3$}\label{sec:pureadsextsur}

\begin{figure}[h]
    \centering
    \includegraphics[scale=0.5]{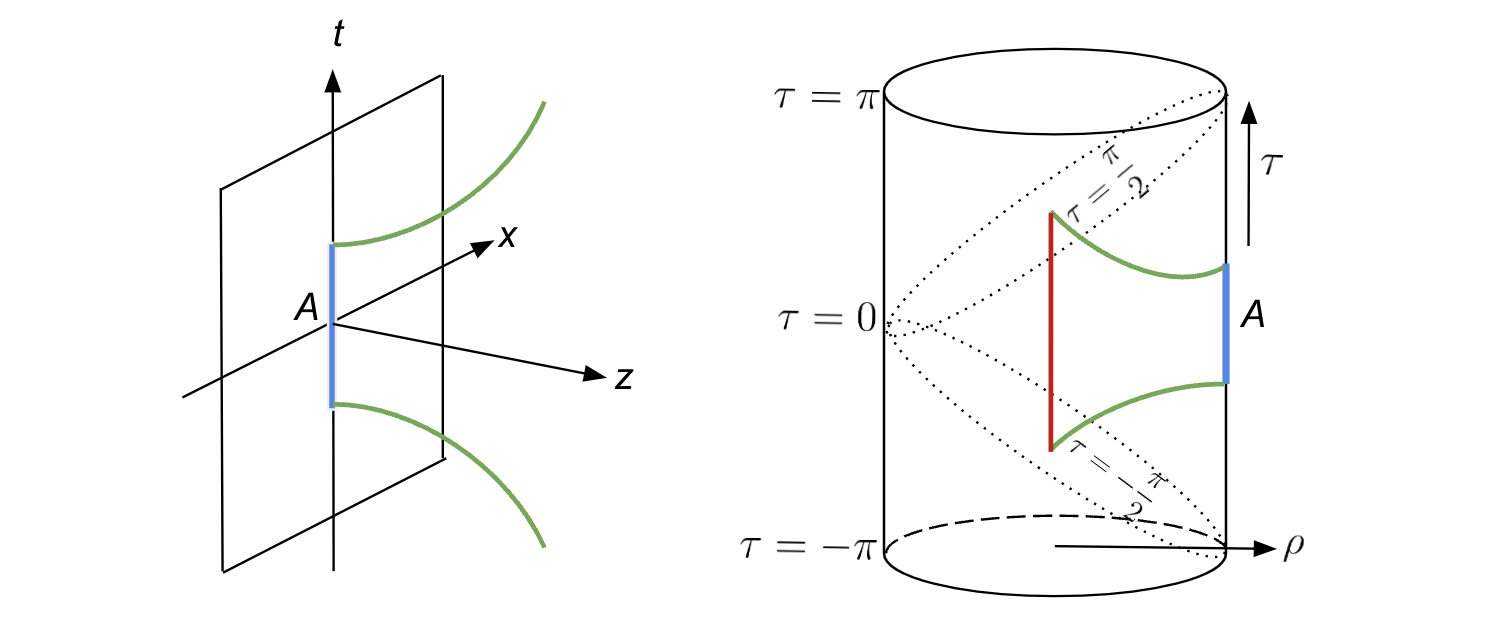}
    \caption{The union of the spacelike (green) and timelike (red) surfaces constitutes the extremal surface for the (blue) timelike subregion $A$.}
    \label{fig:pure}
\end{figure}

We can take the CFT$_2$ in the infinite two dimensional plane to be dual to the Poincar\'{e} AdS$_3$ with metric 
\begin{equation}
    ds^2 = \frac{l^2}{z^2}(dz^2 - dt^2 + dx^2)\,.
\label{201}   
\end{equation}
$l$ and $z$ here are the usual AdS radius and the bulk radial coordinates respectively. If we have a timelike subregion $A$ in CFT$_2$  with endpoints \(t= \pm T_0\) at \(x=0\) plane, then the associated extremal surface takes the shape depicted in figure \ref{fig:pure} (see \cite{Doi:2023zaf} for the details of the derivation using extremization procedure).\footnote{As we will see shortly, the symmetric nature of the depicted extremal surface is due to the symmetric alignment of the timelike region $A$ around $t=0$ slice.}
As we have already seen in (\ref{D}), we expect the timelike EE in this case to be
\begin{equation}
    S_A^{(T)} = \frac{c}{3} \log \left( \frac{2T_0}{\epsilon} \right) + \frac{i \pi c}{6}\,.
\label{200}    
\end{equation}
It turns out that the area of the green spacelike surface given by the equation 
\begin{equation}
    t = \sqrt{z^2 + T_0^2}\,,
\label{202}  
\end{equation}
computes the real part of the above EE; whereas to obtain the imaginary piece, we need to go to the global patch of AdS$_3$. Indeed, it turns out that in global AdS$_3$ with metric
\begin{equation}
    ds^2 = l^2 [-\cosh^2\rho\, d\tau^2 + d\rho^2 + \sinh^2\rho\, d\theta^2]\,,
\label{204}    
\end{equation} 
the spacelike and the timelike segments join precisely at global times $\tau=\pm \pi/2$, therefore yielding the length of the red segment to be $\tau=\pi$. This in turn computes the imaginary piece going as $\frac{i\pi c}{6}$ \cite{Doi:2022iyj,Doi:2023zaf}. The green spacelike parts happen to follow similar equation as the RT surface, but appropriate for a timelike interval.

\subsubsection{BTZ Blackhole}\label{subsubsec:btzextremal}

As the last topic of our review, we describe the timelike extremal surfaces for the BTZ black hole. The corresponding metric is given by
\begin{equation}
    ds^2 = -\frac{(r^2-r_+^2)}{l^2}dt^2 +\frac{l^2}{(r^2-r_+^2)}dr^2 + r^2 d\phi^2\,,
    \label{eq:btzmet}
\end{equation}
where $r_+$ is the horizon radius. Moreover, $\beta = \frac{2\pi l^2}{r_+}$ denotes the inverse temperature in this case.

\begin{figure}[h]
    \centering
    \includegraphics[scale=0.5]{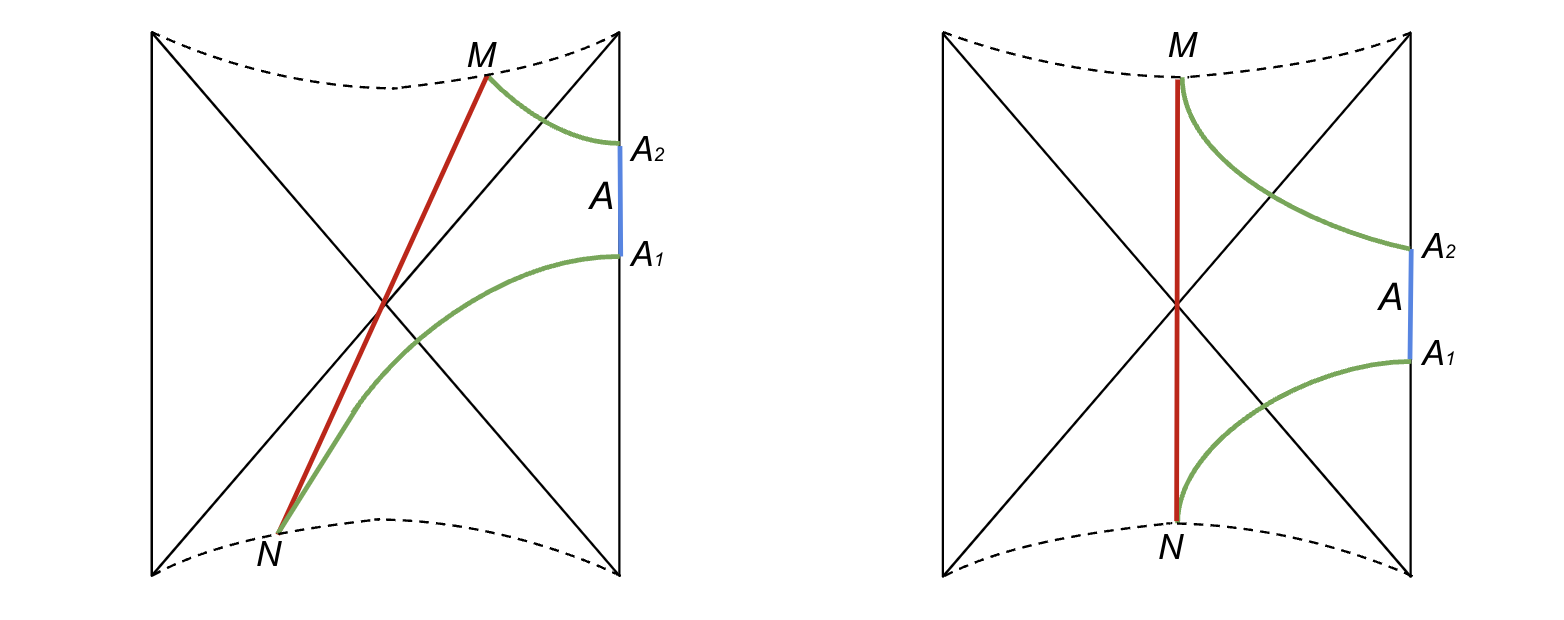}
    \caption{The (left) right panel surfaces respectively depict the extremal surfaces for a timelike subregion which is (not) symmetric around boundary $t=0$.}
    \label{fig:HEE_02}
\end{figure}

Our goal will be to compute the timelike EE of a region $A:=[t=T_1, t=T_2]$  on the right boundary with $T_1<T_2$. Using kruskal coordinates \cite{Doi:2023zaf}, in which the endpoints are denoted by $(v_1,u_1)=(a_1,-\frac{1}{a_1})$ and $(v_2, u_2) = (a_2, -\frac{1}{a_2})$ (where $a_i = e^\frac{r_+ T_i}{l^2}$), we have the following shape of the extremal surface as depicted in figure \ref{fig:HEE_02}.\footnote{The Kruskal coordinates at the right boundary (at $x=0$ slice) is written in terms of Rindler time coordinate as $v=e^{\frac{r_+ t}{l^2}}$ and $u=-e^{-\frac{r_+ t}{l^2}}$.}

The extremal surfaces depicted in figure \ref{fig:HEE_02} consist of two spacelike geodesics $A_2M$ and $A_1N$, which extend between the right boundary and future/ past singularity, and a timelike geodesic $MN$ which connects the two spacelike RT surfaces. For a symmetrically placed (around $t=0$) timelike interval, we have symmetric orientation of the extremal surface. Given these extremal surfaces, it was rigorously shown \cite{Doi:2023zaf} that one obtains the well-known expression of the timelike entanglement entropy \eqref{J}
\begin{equation}
    S_A^{(T)} = \frac{c}{3}\log \Bigg( \frac{\beta}{\pi \epsilon}\sinh{\Bigg(\frac{\pi}{\beta}(T_2-T_1)\Bigg)}\Bigg) + \frac{i\pi c}{6} \label{}\,.
\end{equation}
As we will see later, this interesting fact that the corresponding extremal surfaces naturally probe the geometry beyond the horizon, is what enables one to reconstruct bulk fields in regions that are not causally connected to either of the CFTs separately. 

\section{Timelike modular Hamiltonian}\label{sect:intro2tlEH}

We now turn to the study of (timelike) modular Hamiltonians associated to the timelike EE. As we mentioned in the introduction, modular Hamiltonians are natural objects to be used in the context of bulk reconstruction. For spherical spacelike subregions in holographic CFTs they implement a geometric flow of bulk operators within the corresponding bulk entanglement wedges \cite{Bisognano:1976za,Casini:2011kv}. This is why it has already been used extensively in the context of tracking bulk fields for both the pure AdS and AdS black hole backgrounds \cite{Jafferis:2015del,Dong:2016eik,Kim:2016ipt,Faulkner:2017vdd,Kabat:2017mun,Faulkner:2018faa}, including fields inside black hole horizons \cite{Bao:2019hwq,Jafferis:2020ora,deBoer:2022zps}.

We \emph{define} the timelike modular Hamiltonian $H_A^{(T)}$ in a manner similar to how we have discussed the derivation of the timelike entanglement entropy, starting from the spacelike one (as in section \ref{sec:wickee}). For example, we can employ the Wick rotation given by \eqref{2} and due to the system of our interests, we will choose a starting set-up with a spacelike interval in a given CFT. We will illustrate this with two simple examples of CFTs.

There have been some discussions on the associated Hamiltonians corresponding to pseudo-entropies in \cite{Nakata:2020luh,Doi:2022iyj}. For dS/CFT pseudo-entropy, the Hamiltonian related to the non-hermitian transition matrix is hermitian, whereas our definition for timelike modular Hamiltonian for Lorentzian CFTs will be non-hermitian, giving rise to a non-hermitian density matrix for a timelike interval. Indeed, upon a Wick rotation in the Lorentzian CFT, we will find a Hamiltonian with both real and imaginary pieces; whereas in the context of dS/CFT, they will be purely real (discussed later in section \ref{sect:dS}).

\subsection{Timelike interval in CFT$_2$ on a periodic space}\label{subsec:tlmhps}

Given that we are interested in figuring out the timelike modular Hamiltonian for a CFT$_2$ on a periodic space (with period $L$), our natural starting point would be a spacelike subregion for a CFT at finite temperature (with inverse temperature $\beta$). We already know the usual modular Hamiltonian for the latter case \cite{Cardy:2016fqc}. It is given by\footnote{For subsections \ref{subsec:tlmhps} and \ref{subsec:tlmhft} alone, the coordinate $z$ will denote Euclidean boundary lightcone coordinate. It is not to be confused with the Poincar\'{e} radial coordinate used elsewhere in this paper.} 
\begin{equation}\label{eq:slmh}
    H_{A}=2\beta\int_{A}\frac{\sinh{\left(\frac{\pi(R-z)}{\beta}\right)}\sinh{\left(\frac{\pi(R+z)}{\beta}\right)}}{\sinh{\frac{2\pi R}{\beta}}}T_{zz}(z) \,dz+\text{anti-chiral part} \,.
\end{equation}
This is essentially obtained by implementing the conformal map $z\to w=f(z)$ on the Euclidean lightcone coordinates $z$, which maps the subregion stretched between $z=[-R,R]$ to an annulus. The conformal map in this case is given by 
\begin{equation}\label{eq:ct1}
    f(z)= \log \left(\frac{e^{\frac{2\pi z}{\beta}}-e^{-\frac{2\pi R}{\beta}}}{e^{\frac{2\pi R}{\beta}}-e^{\frac{2\pi z}{\beta}}} \right)
\end{equation}
yielding the usual entanglement entropy \eqref{V} of the spacelike, finite temperature case.

We can now therefore implement the Wick rotation \eqref{2} which maps the subregion to stretch from points \(z=iT_0\) to \(z=-iT_0\). The resulting conformal transformation gets mapped to
\begin{equation}\label{eq:ct2}
    f(z)= \log \left(\frac{e^{\frac{2i\pi z}{L}}-e^{-\frac{2i\pi T_0}{L}}}{e^{\frac{2i\pi T_0}{L}}-e^{\frac{2i\pi z}{L}}} \right)\,.
\end{equation}
Clearly, we have obtained \eqref{eq:ct2} from \eqref{eq:ct1} by identifying the $z$ as $iz$, realizing that $R$ must now be replaced by $iT_0$ and simply replacing $\beta$ by $L$. Under this mapping the width of the annulus goes from
\begin{equation}
     W= f(R-\epsilon)-f(-R + \epsilon) = 2\log\left(\frac{\beta }{\pi \epsilon}\sinh{\frac{2\pi R} {\beta}}\right)
\end{equation}
to 
\begin{equation}
    W= f(iT_0 - \epsilon)-f(-iT_0 + \epsilon) = 2\log\left(\frac{L}{\pi \epsilon}\sinh{\frac{2\pi i T_0}{L}}\right) \,,
\end{equation}
and the new entangling points are mapped to \(e^{\frac{2i\pi T_0}{L}}\) and \(e^{\frac{-2i\pi T_0}{L}}\) respectively.

Finally, the timelike modular Hamiltonian takes the following form (suppressing the anti-chiral part; either using the new annulus width obtained above, or by directly implementing the Wick rotation in \eqref{eq:slmh})
\begin{equation}\label{eq:tlmhps}
    H_{A}^{(T)}=2L\int_{A} \frac{\sin{\left(\frac{\pi(T_0-t)}{L}\right)}\sin{\left(\frac{\pi(T_0+t)}{L}\right)}}{\sin{(\frac{2\pi T_0}{L}})}T_{00}(t) \,dt\,,
\end{equation} 
where $L$ is now interpreted as the periodicity of space and $t$ is the Lorentzian time. We have also put $x=0$ above, along which we take the timelike interval. This timelike modular Hamiltonian is easily generalizable to the case where the subregion is asymmetric (i.e. for $A\in [T_1,T_2]$ and \emph{not} $A\in [-T_0,T_0]$), and we can also take $L\to \infty$ limit to obtain the corresponding timelike Hamiltionian for CFT on a plane (see \eqref{8}). The width of the annulus is related to the Renyi and hence the entanglement entropy, which in this case yields
\begin{equation}
    S_A^{(T)}=\frac{c}{3}\log \left( \frac{L }{\pi \epsilon}\sin{\frac{2\pi T_0}{L}}\right) + \frac{i\pi c}{6} \,,
\end{equation} 
and is precisely the timelike entanglement entropy we obtained previously in \eqref{G} or \eqref{Z}. We will be using this modular Hamiltonian above when reconstructing the bulk fields in pure AdS$_3$, where the special features of the extremal surfaces as discussed in section \ref{sec:pureadsextsur} will be clear.

\subsection{Timelike interval in CFT$_2$ at finite temperature}\label{subsec:tlmhft}

We can carry out a similar manipulation in order to find out the timelike modular Hamiltonian for a CFT at finite temperature. In this case, our natural starting point will be a spacelike subregion of size $2R$ in a CFT along a compact, spatial direction (of size $L$), on which we will implement the Wick rotation.

In this case, the conformal map from Euclidean spacetime coordinate $z$ to $w=f(z)$ is given by 
\begin{equation}
    f(z)= \log \left(\frac{e^{\frac{2\pi i z}{L}}-e^{-\frac{2\pi i R}{L}}}{e^{\frac{2\pi i R}{L}}-e^{\frac{2\pi i z}{L}}} \right)\,,
\end{equation}
which maps the entangling points \(z=R\) and \(z=-R\) to \(e^{\frac{2\pi i R}{L}}\) and \(e^{\frac{-2\pi i R}{L}}\) respectively. In this case, the modular Hamiltonian takes the form
\begin{equation}\label{eq:LSLmodham}
    H_{A}=2L\int_{A}\frac{\sin{\left(\frac{\pi(R-z)}{L}\right)}\sin{\left(\frac{\pi(R+z)}{L}\right)}}{\sin{\frac{2\pi R}{L}}}T_{zz}(z) \,dz +\text{anti-chiral part}\,.
\end{equation}
If we now perform the type I Wick rotation discussed before
\begin{equation}
    {L}={-i\beta} \quad\text{and}\quad\epsilon \to -i\epsilon  \,,
\end{equation} 
it provides us with the required timelike modular Hamiltonian for a timelike interval with the subregion size $R$ identified to $T_0$:
\begin{equation}\label{eq:tlmhbtz}
    H_{A}^{(T)}=2\beta\int_{A} \frac{\sinh{\left(\frac{\pi(T_0-z)}{\beta}\right)}\sinh{\left(\frac{\pi(T_0+z)}{\beta}\right)}}{\sinh{(\frac{2\pi T_0}{\beta}})}T_{zz}(z) \,dz+\text{anti-chiral part}\,.
\end{equation} 

Once again, the width of the annulus under the conformal map changes from 
\begin{equation}
     W= f(R - \epsilon)-f(-R + \epsilon) = 2\log\left(\frac{L}{\pi \epsilon}\,{\sin{\bigg(\frac{2\pi R} {L}\bigg)}}\right)  + O(\epsilon) \label{}
\end{equation} 
to the corresponding timelike one, to provide the timelike Renyi entropy as 
\begin{equation}
    S^{(n) (T)}_A=\frac{c}{6}\left(1+\frac{1}{n}\right)\left(\log \left(\frac{\beta}{\pi \epsilon}\sinh{\frac{2\pi T_0}{\beta}}\right) + i\pi \right) \,.
\end{equation} 
For $n\to 1$, we get the timelike entanglement entropy to be \eqref{J}
\begin{equation}
    S_A^{(T)}=\frac{c}{3} \log \left(\frac{\beta}{\pi \epsilon}\sinh{\frac{2\pi T_0}{\beta}}\right) + \frac{i\pi c}{6}\,.
\end{equation} 
Once again, we will make use of this timelike modular Hamiltonian in order to construct bulk fields in BTZ backgrounds.

An analysis of the resulting modular flows due to Hamiltonians \eqref{eq:tlmhps} and \eqref{eq:tlmhbtz} has been deferred to appendix \ref{app:app}. There we have also constructed the corresponding bulk timelike modular Hamiltonians, which will help us understand the geometric flow of bulk operators under the action of such Hamiltonians. We will see that the bulk fields located on the spacelike parts of the corresponding extremal surfaces (i.e. the green parts of the extremal surfaces in figures \ref{fig:pure} and \ref{fig:HEE_02}) remain invariant under the timelike bulk modular flow, whereas the red timelike locations are not invariant. This will help us reconstruct the bulk fields everywhere within the bulk using the timelike modular Hamiltonians, to which we turn next.

\section{Revisiting AdS$_3$ Poincar\'{e} reconstruction\label{sect:poincareoutside}}

Armed with the knowledge of timelike modular Hamiltonians in the above spacetimes, we are now ready to formulate the extrapolate bulk dictionary closely following the techniques of \cite{Kabat:2017mun}. As mentioned briefly in the introduction, their method utilized the fact that the spacelike RT surface is invariant under the usual (spacelike, extended) modular Hamiltonian. Therefore, any local, free bulk scalar field $\Phi(\gamma)$ located on the RT surface $\gamma$ must commute with the corresponding modular Hamiltonian. In particular, for AdS$_3$, this implies that any bulk scalar sitting at the crossing point of two intersecting RT surfaces (corresponding to two overlapping subregions at the boundary), must commute with both the respective modular Hamiltonians. Turning this statement to a boundary constraint, we can establish the extrapolate dictionary for free bulk fields. Here, we will refrain from reviewing their work any further, and rather direct the readers to \cite{Kabat:2017mun,Kabat:2018smf,Roy:2018ehv} where this method was used in AdS$_3$ to construct bulk scalars and gauge fields, along with the derivation of the background bulk spacetime.

Our work here can be understood as a timelike version of these earlier works, but as we shall see, our boundary constraints contain far reaching information regarding the bulk spacetimes, compared to the spacelike counterparts.\footnote{This is not really apparent in the example of this section. But when applied for black holes such as BTZ, as we have done in later sections, we will see that the  extremal surfaces associated to timelike intervals probe beyond the horizon, whereas the typical RT surfaces do not (as in figure \ref{fig:HEE_02}). This distinction is important in understanding the utility of the current formalism involving extremal surfaces of timelike intervals.} In our case, in order to find out a local bulk field in Poincar\'{e} AdS$_3$ (outside the Poincar\'{e} horizon) with metric \eqref{201}, we will use the expression of timelike modular Hamiltonian as is given in \eqref{eq:tlmhps} (in the limit $L\to \infty$). It takes the form
\begin{equation}
    H_{A}^{(T)}= 2\pi \int_{A} \frac{(t_2-\xi)(\xi - t_1)}{t_2 - t_1}T_{\xi \xi}(\xi) \,d\xi  + \text{anti-chiral part} \label{8}
\end{equation}
for a timelike subregion $A$ extending from $t=t_1$ to $t_2$ in a constant spatial (in this case $x=0$) slice. $\xi$ and $\bar{\xi}$ are the lightcone coordinates defined as $\xi = t-x$ and $\bar{\xi}=t+x$.\footnote{We found it easier to use these slightly different notations for lightcone coordinates (different from \cite{Kabat:2017mun}), as using them we get similar looking equations as in \cite{Kabat:2017mun}, although we are using timelike intervals instead of spacelike intervals. Note that in appendix \ref{appsec:modflowpureads}, we have used slightly different lightcone coordinates ($\omega,\bar{\omega}$) \eqref{eq:lcother}. However, none of the final expressions in the appendix depend on these variables, so this change of convention doesn't really matter.} We will once again use the extended timelike modular Hamiltonian, which turns out to be a better defined quantity to work with as it is once again a simple extension of the timelike subregion $A$ to all over the spatial slice:
\begin{equation}\label{8ext}
    \tilde{H}_{A}^{(T)}= 2\pi \int_{-\infty}^{\infty} \frac{(t_2-\xi)(\xi - t_1)}{t_2 - t_1}T_{\xi \xi}(\xi) \,d\xi  + \text{anti-chiral part}\,.
\end{equation}
In other words, 
\begin{equation}\label{eq:tlexthmod}
	\tilde{H}^{(T)}_A= H_{A}^{(T)}- H_{A^c}^{(T)}\,,
\end{equation}
with $A^c$ denoting the complimentary region to $A$.\footnote{We will need to work with $\tilde{H}^{(T)}_A$ for technical purposes, as only for such quantities, the integrations run from $-\infty$ to $\infty$ as in \eqref{8ext}. However, for spacelike entanglement in QFT, such extended modular Hamiltonians (negative logarithm of the so-called modular operators) are well-defined according to Tomita-Takesaki theory \cite{Witten:2018lha}, and thus is a natural starting point. It will be interesting to investigate whether there is such a natural algebraic reason for timelike extended modular Hamiltonians as well.}
For future use, we note that \eqref{8ext}, combined with the commutator between the CFT$_2$ stress tensor and a CFT boundary scalar primary $\mathcal{O}$ (with chiral conformal dimension $h=\Delta/2$, and similarly for the anti-chiral part)
\begin{equation}
  2\pi [T_{ww}(w),\mathcal{O}(\xi,\bar{\xi})]=2\pi i \,(h\,\partial_{\xi}\, \delta(\xi-w)\mathcal{O}+  \delta(\xi-w)\partial_{\xi}\mathcal{O})
\end{equation}
yields 
\begin{equation}
    [\tilde{H}_{A}^{(T)}, \mathcal{O}(\xi,\bar{\xi})]= \frac{2\pi i}{(t_2-t_1)}\left(\Delta(\bar{\xi}-\xi)-t_1t_2(\partial_{\xi}-\partial_{\bar{\xi}})+(t_1+t_2)(\xi \partial_{\xi}-\bar{\xi}\partial_{\bar{\xi}})+\bar{\xi}^2\partial_{\bar{\xi}}-\xi^2 \partial_{\xi} \right)\mathcal{O}(\xi,\bar{\xi})\,. \label{12}
\end{equation}

In the appendix section \ref{appsec:modflowpureads}, around \eqref{eq:slinv}, we have already checked that the timelike modular Hamiltonian doesn't flow the spacelike part of the extremal surface corresponding to a timelike interval. We can test this explicitly by considering the commutator between a bulk field sitting at this spacelike part of the slice with the timelike modular Hamiltonian given in \eqref{8ext}. From the HKLL formalism, we know that the bulk scalar field $\Phi$ at a bulk point $(x,t,z)$ in Poincar\'{e} coordinates is given by \cite{Hamilton:2006fh} 
\begin{equation}
    \Phi(x, t, z)=\frac{(\Delta-1)}{\pi} \int_{z^2-y'^2-t'^2>0} \left(\frac{z^2-y'^2-t'^2 }{z}\right)^{\Delta-2}\mathcal{O}(t+t', x+iy')\,. \label{13}
\end{equation}
Here we have used the usual complexified coordinate representation (bulk spatial coordinates extended to take complex values), in which the reconstruction formula becomes manifestly holographic. For our purposes here, we consider the bulk field at a constant spatial plane ($x=0$ for specificity), in which case it can be written as 
\begin{equation}
    \Phi(x=0,t,z)=\frac{(\Delta-1)}{\pi} \int_{z^2-y'^2-t'^2>0} \left(\frac{z^2-y'^2-t'^2 }{z}\right)^{\Delta-2}e^{t'\frac{d}{\;dt}}\mathcal{O}(t,iy')\,.
    \label{13}
\end{equation}
With \(\ (\xi=t-iy'\;,\; \bar{\xi}=t+iy')\) (as per our definition earlier), we can evaluate the following commutator as 
\[[\tilde{H}_{A}^{(T)}, \Phi(x=0,t,z)] = \frac{2i(\Delta-1)}{(t_2-t_1)}\int_{z^2-y'^2-t'^2>0} \left(\frac{z^2-y'^2-t'^2 }{z}\right)^{\Delta-2}e^{t'(\frac{d}{\;d\xi}+\frac{d}{d\bar{\xi}})}\]\ \begin{equation}
     \left(\Delta(\bar{\xi}-\xi)-t_1t_2(\partial_{\xi}-\partial_{\bar{\xi}})+(t_1+t_2)(\xi \partial_{\xi}-\bar{\xi}\partial_{\bar{\xi}})+\bar{\xi}^2\partial_{\bar{\xi}}-\xi^2 \partial_{\xi} \right)\mathcal{O}(\xi,\bar{\xi})\,. \label{14}   
 \end{equation} 
Following \cite{Kabat:2017mun} closely and introducing new variables \((q = \xi+t'\,,\, p= \bar{\xi}+t')\), we find
\[ [\tilde{H}_{A}^{(T)}, \Phi(x=0,t,z)] = \frac{2i(\Delta-1)}{(t_2-t_1)}\int_{z^2-(p-t)(q-t)>0} \left(\frac{z^2-(p-t)(q-t)}{z}\right)^{\Delta-2}\]
\begin{equation}
    \left(\Delta(p-q)-t_1t_2(\partial_{q}-\partial_{p})+(t_1+t_2)(q \partial_{q}-p\partial_{p})+p^2\partial_{p}-q^2 \partial_{q} \right)\mathcal{O}(q,p)\,. \label{15}
\end{equation}
Performing some integrations by parts, we find that in order for \([\tilde{H}_{A}^{(T)}, \Phi]=0\), we need
\begin{equation}
    z^2-t_1t_2+(t_1+t_2)t-t^2=0\,. \label{16}
\end{equation} 
This is nothing but the geodesic equation for the spacelike part of the extremal surface corresponding to the timelike subregion. Indeed, when the subregion is symmetric with $(t_1 = -T_0\, ,\, t_2 = T_0)$, \eqref{16} becomes
\begin{equation}
	t^2 = z^2 + T_0^2\,,
\end{equation}
which is what we mentioned in \eqref{202} before.

\subsection{AdS$_3$ Poincar\'{e} reconstruction using $\tilde{H}^{(T)}$}

\begin{figure}[h]
    \centering
    \includegraphics[scale=0.5]{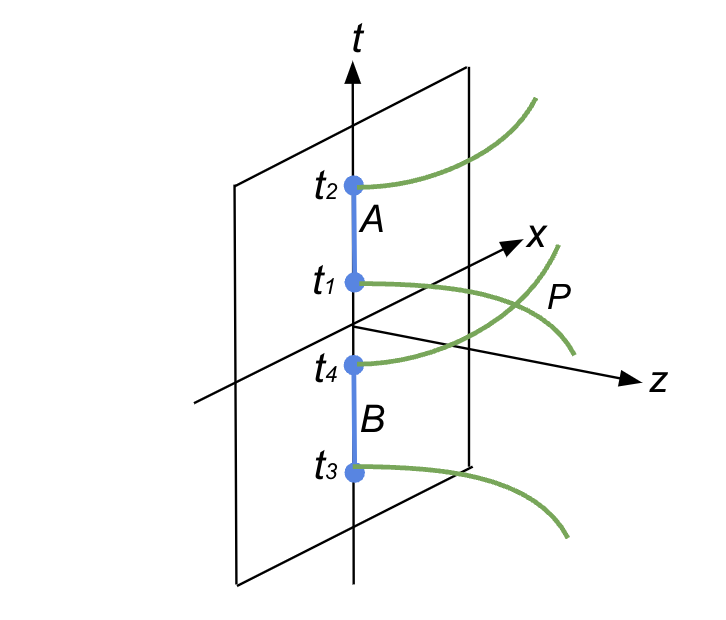}
    \caption{Two non-intersecting  timelike subsystems $A$ and $B$ whose spacelike parts of the corresponding extremal surfaces intersect at a bulk point $P$. }
    \label{fig:Bulk_01}
\end{figure}

We can now revert the observations we made in the previous part of this section, and make it a boundary constraint involving $\tilde{H}^{(T)}$ to re-derive the HKLL scalar fields. In particular, following \cite{Kabat:2017mun}, we start by considering two non-overlapping timelike subregions (as in figure \ref{fig:Bulk_01}. This is unlike what happens in the boundary for the spacelike case), which are extending between \((t_1,t_2)\) and \((t_3,t_4)\) respectively.
The intervals are such that the corresponding extremal surfaces, especially their spacelike parts, intersect at a point $P$ deep inside the bulk.

Given it's a purely boundary reconstruction, we pick an ansatz for our bulk field as a non-local boundary operator and located on the boundary spatial $x=0$ slice, and a generic point $P$ as given below\footnote{$p,q$ here are defined as $p=T+t'-iy'$ and $q=T+t'+iy'$ consistent with our previous definitions.}
\begin{equation}
    \Phi(P)=\int dt'dy'\,g(q,p)\,\mathcal{O}(q,p)\,. \label{18}
\end{equation}
Here \(g(q,p)\) is the smearing function we want to evaluate. If we impose that the above field commutes with the timelike modular Hamiltonians of both these timelike subregions, in other words (with some minor changes. The subscript $(ij)$ in $\tilde{H}^{(T)}$ denotes the subregion in question)
\begin{equation}
    (t_2-t_1)[\tilde{H}^{(T)}_{12},\Phi]-(t_4-t_3)[\tilde{H}^{(T)}_{34},\Phi]=0\,, \label{19}
\end{equation}
then we can reproduce the corresponding HKLL bulk field located at some time coordinate given by $t=t_*$ (defined below), and at an \emph{emergent} bulk direction $z$ which is a certain function of $t_*$ and the endpoints (and on the boundary spatial $x=0$ slice). The relation that $z$ will satisfy, will essentially indicate that the bulk field is on the spacelike part of the extremal surface associated with the timelike interval. Indeed, we see that \eqref{19} is alternatively written as
\begin{equation}
    \int dq\,dp\,g(p,q) \left[(t_3t_4-t_1t_2)(\partial_{q}-\partial_{p})+(t_1+t_2-t_3-t_4)(q \partial_{q}-p\partial_{p})\right]\mathcal{O}(q,p) =0\,. \label{20}
\end{equation}
Carrying out some integrations by parts, we arrive at 
\begin{equation}
    \left[(t_*-q)\partial_{q}-(t_*-p)\partial_{p}\right]\,g(q,p)=0\,, \label{21}
\end{equation}
 where \(t_*=\frac{t_1t_2-t_3t_4}{t_1+t_2-t_3-t_4}\). 
This is solved by a general function $f$ of the form
\begin{equation}
    g(q,p)=f((q-t_*)(p-t_*))\,. \label{22}
\end{equation}
If we use this form in e.g. $[\tilde{H}^{(T)}_{12},\Phi]=0$\,, we can find out the explicit form of the function $g$ as 
    \begin{equation}
    g(q,p)=c_{\Delta} \left(z^2-(q-t_*)(p-t_*) \right)^{\Delta-2}
    \label{23}
\end{equation} 
with 
\begin{equation}
    z^2=t_1t_2-(t_1+t_2)t_*+t_*^2\,. \label{24}
\end{equation} 
The coefficient $c_\Delta$ can be found in exactly the same way (using normalizable mode condition) as in \cite{Kabat:2017mun}. Using the equation of the spacelike geodesic \eqref{16} and realizing that $t_*$ is the value of the $t$ coordinate at the intersecting point $P$ in figure \ref{fig:Bulk_01}, we conclude that the local bulk field is located at point $P(z,t_*)$, with $z$ given by \eqref{24} and have the form
\begin{equation}
 \Phi(P)=   \Phi(z,t_*)=c_{\Delta}\int_{z^2-(q-t_*)(p-t_*)>0} dt'dy' \left(z^2-(q-t_*)(p-t_*)\right)^{\Delta-2}\mathcal{O}(t_*+t',iy')\,. \label{25}
\end{equation}
Unsurprisingly, this is of course the correct form of the local bulk field \`{a} la HKLL, this time derived entirely from boundary constraints \eqref{19}. Therefore, we can conclude that the boundary constraints \eqref{19} gives rise to the HKLL prescription along with the bulk emergent direction $z$, which satisfies equation of the extremal surface appropriate for this case.

\section{Outside BTZ horizon\label{sect:btzout}}

We now turn our attention to a construction similar to the last section, but now in the background of BTZ black holes (once again, the case for Rindler should follow the same way). Usually, for local bulk fields placed outside the black hole horizon, it is sufficient to know the usual (for spacelike subregions) entanglement entropy and the associated modular Hamiltonians. But, as we will see, for bulk fields inside the horizon, the timelike entanglement seems to be a natural starting point (to be discussed in the next section \ref{sect:inside}).\footnote{Although it may be possible to use the entanglement for a union of the left and right CFT's spacelike intervals (the entangled pair of the two CFTs that give rise to the bulk black hole state) \cite{Hartman:2013qma}, to reconstruct subhorizon regions. We have not pursued that direction here, although we will have some comments in the conclusion section.}

This section will be devoted to bulk reconstruction outside the BTZ horizon. We have already checked in sections \ref{appsec:modflowbtzout} and \ref{appsec:modflowbtzin} that the spacelike parts of the extremal surface associated to a timelike subregion (located in one of the boundaries dual to BTZ black hole), is invariant under the flow of the corresponding timelike modular Hamiltonian. We will once again check this explicitly by using the known form of the HKLL bulk fields, and later on, revert this line of argument to \emph{re-derive} the HKLL formula.

\begin{figure}[h]
    \centering
    \includegraphics[scale=0.5]{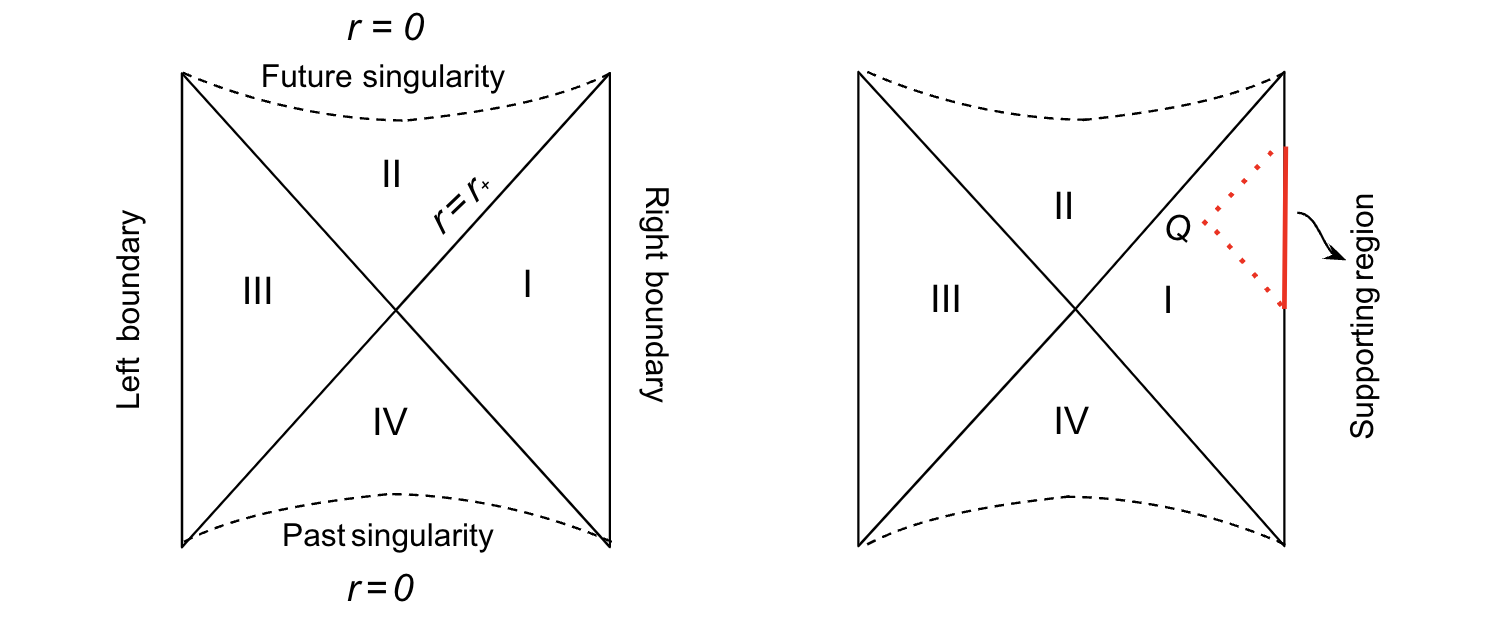}
    \caption{Left: BTZ black hole made up of an entangled pair of the left and right CFTs. Right: HKLL prescription for a bulk scalar located at a point $Q$ outside the BTZ black hole.}
    \label{fig:BTZ_01}
\end{figure}

In other words, our starting point would be a CFT state in the thermofield double form made up of two maximally entangled CFTs
\begin{equation}\label{eq:TFD}
	|0\rangle_{CFT}=\frac{1}{\sqrt{Z}}\sum_i e^{-\beta E_i/2}\,|i\rangle\otimes|i\rangle\,.
\end{equation}
These CFTs would have to satisfy all the holographic properties for it to be dual to a BTZ state in the bulk \cite{Maldacena:2001kr}. Therefore, for this state, we are asking whether considering timelike subregions in either of the CFTs (or both) can lead to a known form of bulk fields either outside or inside of the horizons of the dual geometry. 

Naturally, we will work in the boundary coordinates that appear at the boundary of the BTZ metric given by \eqref{eq:btzmet}. For this background, the associated spacetime diagram and the HKLL support regions are given in figure \ref{fig:BTZ_01}. Here we will use a slightly different version of the finite temperature, timelike modular Hamiltonian than what we wrote down in \eqref{eq:tlmhbtz}. In particular, here we have used the fact that the temperature corresponds to the temperature of the BTZ black hole (given in subsection \ref{subsubsec:btzextremal}), and have written the Hamiltonian in the boundary lightcone coordinates (defining $\xi=t-l\phi$ and $\bar{\xi}=t+l\phi$). We then obtain (going to the corresponding extended timelike modular Hamiltonian given in \eqref{eq:tlexthmod})
\begin{equation}
    \tilde{H}_A^{(T)}=\tilde{c} \int_{-\infty}^{\infty}d\xi\, \left(\cosh\left(\frac{r_+ T_0}{l^2}\right)-\cosh\left(\frac{r_+ \xi}{l^2}\right) \right)T_{\xi{\xi}} + \text{anti-chiral part} \label{26}
\end{equation} 
with
\[ \tilde{c}=\frac{2\pi l^2}{r_+ \sinh\left(\frac{r_+ T_0}{l^2}\right)}\,.\]
For future use we will also need the commutator (with a scalar primary $\mathcal{O}$ of chiral conformal dimension $(h,\bar{h})$)
\begin{eqnarray}\label{eq:btzhextOcomm}
	\left[\tilde{H}_A^{(T)}, \mathcal{O}\right]=\Tilde{c}\,\Bigg(\frac{r_+ h}{l^2}\left(\sinh\left(\frac{r_+\bar{\xi}}{l^2}\right)-\sinh\left(\frac{r_+\xi}{l^2}\right)\right)+\left(\cosh\left(\frac{r_+T_0}{l^2}\right) -\cosh\left(\frac{r_+ \xi}{l^2}\right)\right)\partial_{\xi}\nonumber\\
	-\left(\cosh\left(\frac{r_+T_0}{l^2}\right)-\cosh\left(\frac{r_+\bar{\xi}}{l^2}\right)\right)\partial_{\bar{\xi}}\Bigg)\mathcal{O}(\xi,\bar{\xi})\nonumber\\
\end{eqnarray}

On the other hand, the HKLL scalar field in this background takes the form \cite{Hamilton:2006fh}
\begin{equation}
    \Phi(\phi,t,r) = c_\Delta \int_{spacelike}dy\,dx\,\left[\frac{r}{r_+} \left(\cos{y} \mp \sqrt{1-\frac{r_+^2}{r^2}} \cosh{x} \right)\right]^{\Delta-2}\,\mathcal{O}(\phi + \frac{ily}{r_+}y,t+\frac{l^2 x}{r_+})\,.\label{27}
\end{equation}
Here $\mp$ denotes whether the bulk field in question is on the region I or region III of figure \ref{fig:BTZ_01}, and $c_{\Delta}=\frac{(\Delta-1)(2)^{\Delta-2}l^\Delta}{\pi}$.\footnote{It seems to us that this overall coefficient was wrong in \cite{Hamilton:2006fh}.} For what follows, we redefine the lightcone coordinates by 
\begin{equation}
    \xi=t-\frac{il^2y}{r_+} \;,\; \bar{\xi}=t+\frac{il^2y}{r_+}\,, \label{29}
\end{equation} 
and consequently define 
\begin{equation}
    q=\xi+\frac{l^2 x}{r_+} \;,\;p=\bar{\xi}+\frac{l^2 x}{r_+}\,. \label{30}
\end{equation}
Upon using \eqref{eq:btzhextOcomm} we then obtain
\begin{align}
	&\left[\tilde{H}_{A}^{(T)},\Phi(\phi=0,t,r)\right]\nonumber\\
	&=c_{\Delta}\,\tilde{c}\,\left(\frac{r}{r_+}\right)^{\Delta-2} \int_{spacelike} dq\,dp\, \left(\cosh\left(\frac{r_+(p-q)}{2l^2}\right)-\sqrt{1-\frac{r_+^2}{r^2}} \cosh\left(\frac{r_+(p+q-2t)}{2l^2}\right)\right)^{\Delta-2}\nonumber\\
	& \Bigg[\frac{r_+ \Delta}{2l^2}\left(\sinh\left(\frac{r_+ p}{l^2}\right)-\sinh\left(\frac{r_+ q}{l^2}\right)\right)-\left(\cosh\left(\frac{r_+T_0}{l^2}\right)-\cosh\left(\frac{r_+ q}{l^2}\right)\right)\partial_{q}\nonumber\\
	&+\left(\cosh\left(\frac{r_+T_0}{l^2}\right)-\cosh\left(\frac{r_+p}{l^2}\right)\right)\partial_{p}\Bigg]\ \mathcal{O}(q,p)\,. \label{33}
\end{align}
A brief analysis shows that for the above commutator to be zero, we must have 
\begin{equation}
    \sqrt{1-\frac{r_+^2}{r^2}}=\frac{\cosh({\frac{r_+T_0}{l^2}})}{\cosh(\frac{r_+t}{l^2})}\,, \label{34} 
\end{equation}
which is nothing but the spacelike part of the extremal surface in BTZ background as we mentioned in \eqref{eq:slparttlbtz}.

\subsection{Outside BTZ using $\tilde{H}^{(T)}$}

\begin{figure}[h]
    \centering
    \includegraphics[scale=0.5]{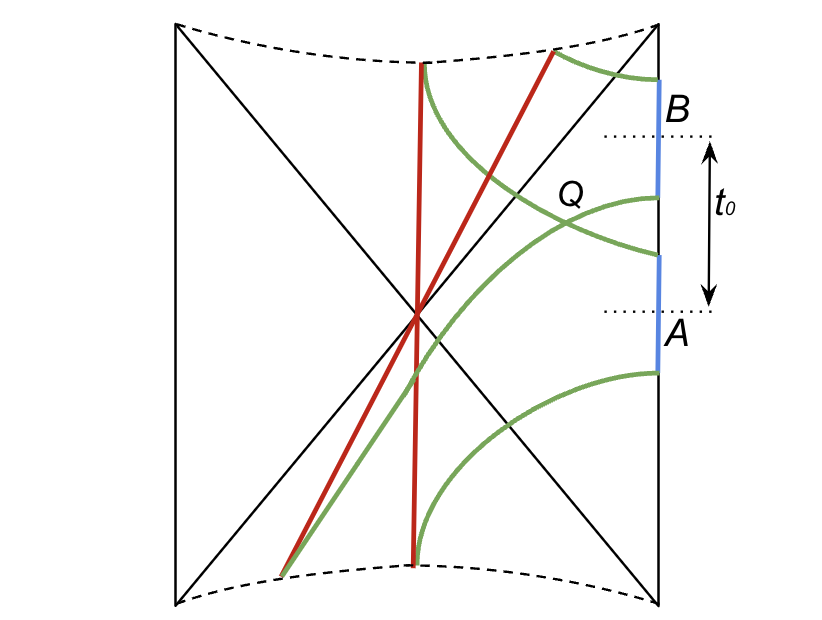}
    \caption{Two timelike subsystems $A$ and $B$ with interval sizes $2T_A$ and $2T_B$ with an intersection of the spacelike parts of their corresponding extremal surfaces (the green lines) at point $Q$. Choosing the center of the interval $A$ at boundary time $t=0$, we find that $Q$ is located outside the BTZ horizon.}
    \label{fig:BTZ_02}
\end{figure}

We are now ready to reverse the argument. The set-up we have in mind is figure \ref{fig:BTZ_02}, with the distance between the intervals being $t_0$. The boundary constraint that is naturally imposed is given by
\begin{equation}\label{eq:bdrycondoutbtz}
    [\tilde{H}_{B}^{(T)},\Phi]=[\tilde{H}_{A}^{(T)},\Phi]=0\,.
\end{equation} 

In a given boundary lightcone coordinate $(\xi,\bar{\xi})$, if the timelike modular Hamiltonian of region $A$ is taken to have a commutator \eqref{eq:btzhextOcomm} with the boundary scalar primary, then for another interval which is $t_0$ time away, we can write the corresponding commutator as\footnote{Here the parameters $r_+$ and $l$ in $\tilde{H}_B^{(T)}$ are just some length scales of the boundary CFT, which has its origin on the temperature of the CFT.}
\begin{align}\label{eq:btzhextOcomm2}
	&\left[\tilde{H}_B^{(T)}, \mathcal{O}\right]=\Tilde{c}\,\Bigg(\frac{r_+ h}{l^2}\left(\sinh\left(\frac{r_+(\bar{\xi}-t_0)}{l^2}\right)-\sinh\left(\frac{r_+(\xi-t_0)}{l^2}\right)\right)\nonumber\\
	&+\left(\cosh\left(\frac{r_+T_B}{l^2}\right) -\cosh\left(\frac{r_+ (\xi-t_0)}{l^2}\right)\right)\partial_{\xi}
	-\left(\cosh\left(\frac{r_+T_B}{l^2}\right)-\cosh\left(\frac{r_+(\bar{\xi}-t_0)}{l^2}\right)\right)\partial_{\bar{\xi}}\Bigg)\mathcal{O}(\xi,\bar{\xi})\,.
\end{align}
Here $2T_B$ is the length of the time interval for region $B$, which may or may not be equal to the interval size $2T_A$ for the region $A$. 

Sure enough, the above commutators in \eqref{eq:bdrycondoutbtz} can be solved using the method of characteristics in a manner closely resembling \cite{Kabat:2017mun}. For this reason, and this construction being for a region outside the black hole (which was already considered in \cite{Kabat:2017mun}),  we have moved the details to appendix \ref{app:outsidebtztechnical}. Indeed we can rigorously show that in the timelike case as well, we can recover \eqref{27} for the bulk field, when the field $\Phi$ is located precisely at a point given by the intersection point $Q$ of the two spacelike geodesics as depicted in figure \ref{fig:BTZ_02}.

\section{Inside the BTZ black hole\label{sect:inside}}

We now turn our attention to bulk fields inside BTZ horizon. As was discussed in \cite{Hamilton:2006az,Hamilton:2006fh}, from bulk perspective, we necessarily require boundary operators from both the left and right CFTs in order to construct such fields. This is related to the boundary entangled structure of the bulk state, which in this case is the thermofield double state \cite{Maldacena:2001kr}. However, such a reconstruction is not readily doable using the spacelike entanglement Hamiltonian even with the knowledge of the entire left and right CFTs separately. The knowledge of spacelike entanglement in any given CFT will only be able to get us up to the horizon (unless we have special cases, where entanglement wedge sees the entanglement shadow regions \cite{Rangamani:2016dms}, not probed by RT surfaces, or unless we know the spacelike entanglement for a \emph{union} of left and right CFTs \cite{Hartman:2013qma}. While the latter approach is in principle possible, we don't have much to say about it in this paper). We will however see, that if one starts with timelike entanglement, and considers two timelike subregions \emph{separately} on both left and right CFTs, the interior reconstruction follows straightforwardly just like it is done for exterior regions. In other words, one doesn't need to go through the complications that arise when one considers the spacelike entanglement of union of two intervals (which usually appear in multipartite systems).

This section is divided into two subsections: in the first, we show that the HKLL \emph{interior} bulk fields are invariant under timelike modular Hamiltonian, if located on the spacelike parts of the extremal surfaces. We have already found this out by an explicit reconstruction of bulk timelike modular Hamiltonian in appendix \ref{appsec:modflowbtzin}. And on the second, we will revert this argument to derive a bulk field for the interior.

\subsection{Timelike modular flow invariance of interior HKLL fields\label{subsec:inside}}

\begin{figure}[h]
    \centering
    \includegraphics[scale=0.5]{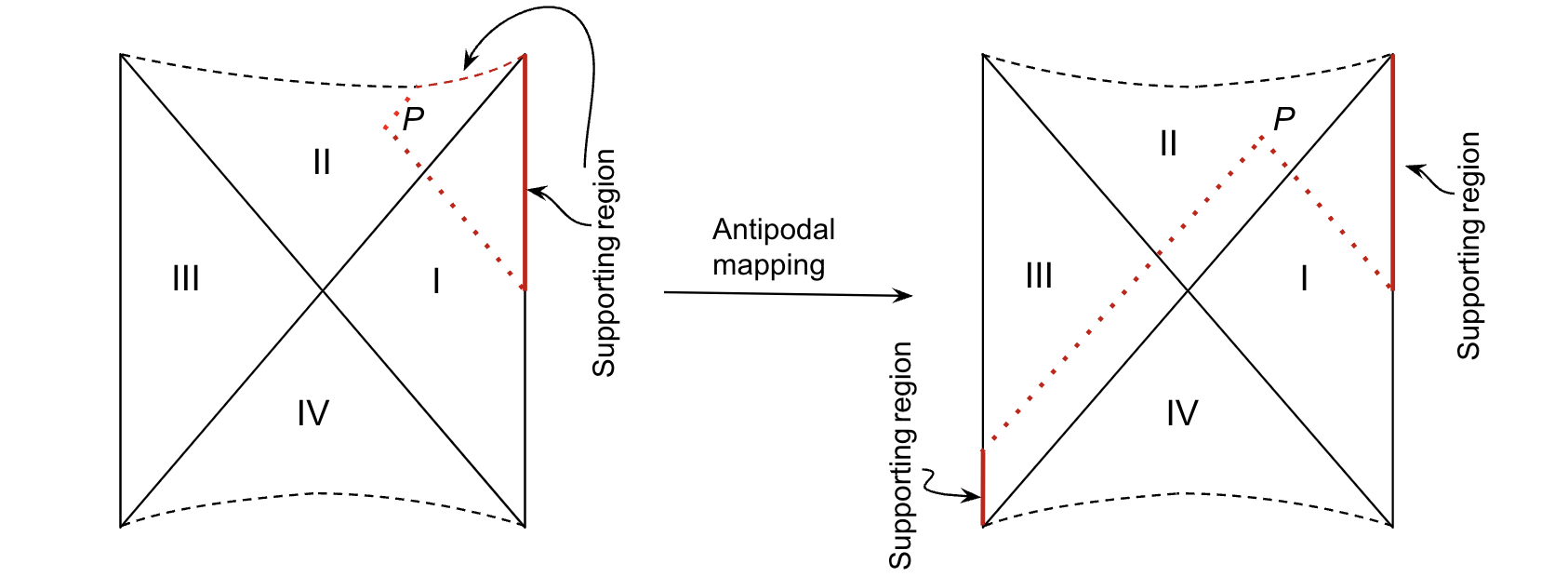}
    \caption{HKLL reconstruction for interior fields at point $P$ requiring support from both boundaries. Supporting region from future singularity can be projected to the left boundary region via an antipodal mapping in this example.}
    \label{fig:BTZ_03}
\end{figure}

We start with the same modular Hamiltonian for timelike subregions in a BTZ Black Hole that we have already started out with in \eqref{26}. The HKLL bulk fields living inside the BTZ black hole can be reconstructed in a way described in figure \ref{fig:BTZ_03}.
The explicit expression is \cite{Hamilton:2006fh} 
\begin{align}
    \Phi(\phi,t,r)&= c_{\Delta}\Big[ \int_{\sigma>0} dy\,dx\, \left(\frac{\sigma_-}{r'}\right)^{\Delta-2} \,\mathcal{O}^{R}\left(\phi + \frac{ily}{r_+},t+\frac{l^2 x}{r_+}\right)\nonumber\\
   &+\int_{\sigma<0}dy'\,dx'\, \left(\frac{\sigma_+}{r'}\right)^{\Delta-2} (-1)^{\Delta}\,\mathcal{O}^{L}\left(\phi + \frac{ily'}{r_+},t+\frac{l^2 x'}{r_+}\right)\Big]=\Phi_R+\Phi_L \label{27in}
\end{align} 
with 
\begin{equation}
    \sigma_{\mp}(\hat{t},r,\hat{\phi}|\hat{t}+x,r',\hat{\phi}+iy)= \frac{rr'}{r_+^2} \left(\cos{y} \mp \sqrt{\frac{r_+^2}{r^2}-1}\, \sinh{x} \right)\,.\label{28}
\end{equation}
The unprimed and primed $(x,y)$ coordinates above denote right and left boundary coordinates respectively. Moreover, $\hat{t}=\frac{r_+ t}{l^2}$, $\hat{\phi}=\frac{r_+ \phi}{l}$ and $c_{\Delta}=\frac{(\Delta-1)(2)^{\Delta-2}l^\Delta}{\pi}$ as before. $\sigma$ here is the bulk to bulk AdS covariant distance with the $\mp$ sign denoting points for the right/left Rindler wedge (with boundary operators $\mathcal{O}^{R}$ and $\mathcal{O}^{L}$ respectively). An inherent limit of $r'\to\infty$ has been taken in the above expression. Also, in the integration limit $\sigma>0$ and $\sigma<0$ will basically denote the spacelike support over the right and left boundary respectively. 

We have written the two terms on the right hand side (RHS) of \eqref{27in} as $\Phi_R$ and $\Phi_L $ as they respectively depend on the right and left boundary operators (although neither of them denote any sort of local bulk operator at the point $(\phi,t,r)$). We will now compute their commutator with the corresponding timelike modular Hamiltonian of the \emph{right} boundary, and see if we can reproduce the conclusion that we obtained in appendix \ref{appsec:modflowbtzin} mentioned above.

Defining the lightcone coordinates $(\xi,\bar{\xi})$, and $(q,p)$ like in section \ref{sect:btzout}, we can straightforwardly see that at least for the first term of \eqref{27in} (in the middle equality) we have an expression very similar to \eqref{33}, which is
\begin{align}
	&\left[\tilde{H}_{R}^{(T)},\Phi_R(\phi=0,t,r)\right]\nonumber\\
	&=c_{\Delta}\,\tilde{c}\,\left(\frac{r}{r_+}\right)^{\Delta-2} \int_{spacelike} dq\,dp\, \left(\cosh\left(\frac{r_+(p-q)}{2l^2}\right)-\sqrt{\frac{r_+^2}{r^2}-1} \sinh\left(\frac{r_+(p+q-2t)}{2l^2}\right)\right)^{\Delta-2}\nonumber\\
	& \Bigg[\frac{r_+ \Delta}{2l^2}\left(\sinh\left(\frac{r_+ p}{l^2}\right)-\sinh\left(\frac{r_+ q}{l^2}\right)\right)-\left(\cosh\left(\frac{r_+T_0}{l^2}\right)-\cosh\left(\frac{r_+ q}{l^2}\right)\right)\partial_{q}\nonumber\\
	&+\left(\cosh\left(\frac{r_+T_0}{l^2}\right)-\cosh\left(\frac{r_+p}{l^2}\right)\right)\partial_{p}\Bigg]\ \mathcal{O}^R(q,p)\,. \label{33in}
\end{align}
The only change is in the second term in the integrand above due to the difference between the AdS covariant distance functions \eqref{27} and \eqref{28}.

We can now impose the condition 
\begin{equation}\label{eq:intcomms}
 [\tilde{H}_R^{(T)},\Phi]=0\qquad\implies\qquad  [\tilde{H}_R^{(T)},\Phi_R]+ [\tilde{H}_R^{(T)},\Phi_L]=0\,,
\end{equation}
and evaluate each of the terms separately. Here $\tilde{H}_R^{(T)}$ is the (extended) modular Hamiltonian for a timelike interval at the right boundary for interval size $2T_0$. We can naively expect that the timelike modular Hamiltonian operator of the right boundary wouldn't flow the left boundary operator $\mathcal{O}^L$, thereby making $[\tilde{H}_R^{(T)},\Phi_L]=0$. Then a computation similar to section \ref{sect:btzout} gives an equation of the form
\[
\left(\sqrt{\frac{r_+^2}{r^2}-1}\sinh\left(\frac{r_+t}{l^2}\right)+\cosh\left({\frac{r_+T_0}{l^2}}\right)\right) \left(\int_{\sigma>0} \left(\frac{\sigma_-}{r'}\right)^{\Delta-3}\sinh\left(\frac{r_+(p-q)}{l^2}\right)\mathcal{O}^R(q,p)\right)=0\,.
\] 
This equation is generically satisfied for the following condition 
\begin{equation}
    \sqrt{\frac{r_+^2}{r^2}-1}=-\frac{\cosh({\frac{r_+T_0}{l^2}})}{\sinh(\frac{r_+t}{l^2})}\,. \label{45}
\end{equation}
This is again the spacelike geodesic for a time-like subregion but going inside a BTZ Black hole \eqref{eq:btzext}.

Interestingly, it turns out that there is another way of deriving \eqref{45}. Knowing the entanglement structure of the boundary state \eqref{eq:TFD} (in other words, that BTZ is a thermofield double state), one may analytically continue the left boundary operator $\mathcal{O}^L$ back to the right boundary. This is often achieved by analytically continuing the time coordinates to complex values \cite{Fidkowski:2003nf,Roy:2015pga}. Following this argument, one might be tempted to conclude that  $[\tilde{H}_R^{(T)},\Phi_L]\neq 0$. We will show in appendix \ref{app:insidebtztechnical} below that even when both the commutators of  \eqref{eq:intcomms} are non-zero, \eqref{45} still holds.

\subsection{Interior reconstruction from boundary constraints\label{subsec:inside2}}

\begin{figure}[h]
    \centering
    \includegraphics[scale=0.5]{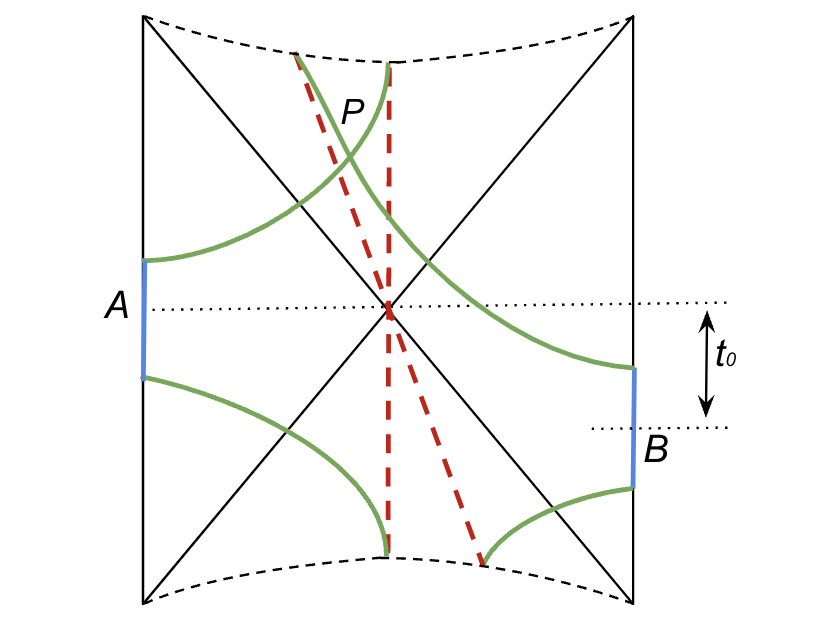}
    \caption{Two timelike subsystems $A$ and $B$ with an intersection of their extremal surfaces at a point $P$ inside the BTZ horizon. Note that we are talking about the spacelike (green) parts. The timelike (red, dashed) parts also intersect, but it is irrelevant for our construction.}
    \label{fig:BTZ_05}
\end{figure}

Our next goal is to construct a local bulk field inside a BTZ black hole. We again start with an asymmetric two-CFT system, so that we have two timelike subsystems at either boundaries, with an intersection point for their corresponding extremal surfaces behind the black hole horizon. Let the size of the left subsystems be \(2T_A\), and the origin of one subregion has an offset by \(t_0\). See figure \ref{fig:BTZ_05}.

The extended Modular Hamiltonian remains the same as before \eqref{26} 
\begin{equation}
    \tilde{H}_A^{(T)}=\tilde{c} \int_{-\infty}^{\infty}d\xi\, \left(\cosh\left(\frac{r_+ T_A}{l^2}\right)-\cosh\left(\frac{r_+ \xi}{l^2}\right) \right)T_{\xi{\xi}} + \text{anti-chiral part}\,, \label{26again}
\end{equation} 
where for $\tilde{H}_B^{(T)}$, we simply have to shift the lightcone coordinates by an appropriate factor of $t_0$. In other words, we have (considering the subsystem $B$ is of size $2T_B$) \eqref{eq:btzhextOcomm2}
\begin{align}
	&[\tilde{H}_B^{(T)},\mathcal{O}]=\Tilde{c}\Biggr(\frac{r_+ h}{l^2}\left(\sinh\left(\frac{r_+(\bar{\xi}-t_0)}{l^2}\right)-\sinh\left(\frac{r_+(\xi-t_0)}{l^2}\right)\right)\nonumber\\
	&+\left(\cosh\left(\frac{r_+T_B}{l^2}\right)-\cosh\left(\frac{r_+ (\xi-t_0)}{l^2}\right)\right)\partial_{\xi}
	-\left(\cosh\left(\frac{r_+T_B}{l^2}\right)-\cosh\left(\frac{r_+(\bar{\xi}-t_0)}{l^2}\right)\right)\partial_{\bar{\xi}}\Biggr) \mathcal{O}(\xi,\bar{\xi})\,.
\end{align}
Even though we are denoting the lightcone coordinates of both boundary's timelike modular Hamiltonians with ($\xi,\bar{\xi}$), note that they define lightcone coordinates of different boundaries.
We once again demand that the boundary constraints are 
\begin{equation}\label{eq:btzinsideconstr}
	[\tilde{H}_A^{(T)},\Phi(P)]=[\tilde{H}_B^{(T)},\Phi(P)]=0\,,
\end{equation}
and solve for the boundary field by making an ansatz like 
\begin{align}\label{eq:btzprimaryansatz}
    \Phi&=\int dq\,dp\,g^R(q,p)\,\mathcal{O}^R(q,p)+\int dq'\,dp'\,g^L(q',p')\,\mathcal{O}^L(q',p')\nonumber\\
    &=\Phi_R+\Phi_L
\end{align}
with $q,p$ defined as in
\begin{equation}
    q=t-\frac{il^2y}{r_+}+\frac{l^2 x}{r_+} \;,\; p=t+\frac{il^2y}{r_+}+\frac{l^2 x}{r_+}\,.
\end{equation}
$(q',p')$ are also defined in the exact same way, but in their case the space and time coordinates are that of the left boundary.
The reason for including conformal operators from both sides of the CFT (rather than just from one CFT as in \eqref{eq:outbtzphiansatz}) is our knowledge that by using $\mathcal{O}^R$ alone, we can't possibly reconstruct the black hole interior, simply because of causality (as is also clear from the computations of the last section). The only other set of operators that we have at our disposal (because we are considering CFT states given by \eqref{eq:TFD}) are the ones from the left boundary, which we are then naturally adding to our ansatz. 
\begin{figure}[h]
    \centering
    \includegraphics[scale=0.5]{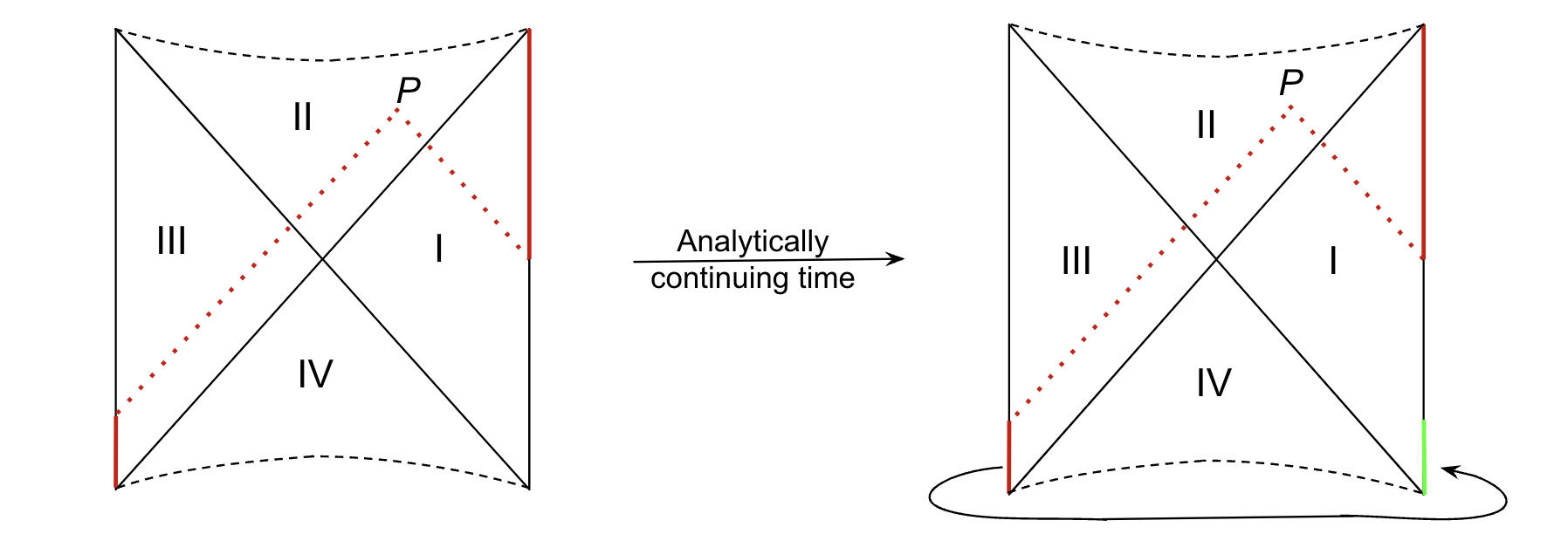}
    \caption{Left: Initially we have the left and right boundary operators $\mathcal{O}^L$ and $\mathcal{O}^R$. Right: We continue the time of $\mathcal{O}^L$ to imaginary values to bring them to the right region owing to the thermofield structure of the CFT state. The point $P$ is where the intersecting point is (as in figure \ref{fig:BTZ_05}).}
    \label{fig:BTZ_INT}
\end{figure}

Working first with the asymmetric interval $B$, we have one of the boundary constraints in \eqref{eq:btzinsideconstr} as
\begin{equation}\label{eq:boundcon0}
    [\tilde{H}_B^{(T)},\Phi_R]+[\tilde{H}_B^{(T)},\Phi_L]=0\,.
\end{equation}
Given $\Phi_L$ is a left boundary operator, we expect it to commute with $\tilde{H}_B^{(T)}$ (similarly, one expects $\Phi_R$ to commute with $\tilde{H}_A^{(T)}$ that we will use later).
However, in \eqref{eq:btzprimaryansatz} one may want to take the left boundary operators to the right boundary (as given by \eqref{eq:kms}), owing to the entangled structure of the CFT state \eqref{eq:TFD} \cite{Roy:2015pga}. In picture, it looks like figure \ref{fig:BTZ_INT}. The expression of $\Phi$ then becomes
\begin{align}
     \Phi&=\int dq\,dp\,g^R(q,p)\,\mathcal{O}^{R}(q,p)+\int dq^{new}\,dp^{new}\,g^L(q^{new},p^{new})\,\mathcal{O}^{R}(q^{new},p^{new})\nonumber\\
     &=\Phi_R+\bar{\Phi}_R\,.
\end{align} 
Here we have denoted the second term above by $\bar{\Phi}_R$ instead of $\Phi_L$ to emphasize that it is now made up of boundary operators residing on the right CFT. Also, $p^{new}$ and $q^{new}$ are defined exactly the same way as in \eqref{42}. Under these changes, the boundary constraint \eqref{eq:boundcon0} becomes
\begin{equation}\label{eq:boundcon}
    [\tilde{H}_B^{(T)},\Phi_R]+[\tilde{H}_B^{(T)},\bar{\Phi}_R]=0\,.
\end{equation}
Even though $\bar{\Phi}_R$ is made up of the operators on the right boundary (the green supported region on the lower right in figure \ref{fig:BTZ_INT}), it still commutes with the operators from the right boundary, such as $\tilde{H}_B^{(T)}$ (a similar story will hold for $\tilde{H}_A^{(T)}$ and $\bar{\Phi}_L$). See e.g. section 3.4 of \cite{Lashkari:2018oke}.
In other words, the individual commutators appearing in \eqref{eq:boundcon} are zero. 
This leads to the following two equations\footnote{Once again, for any variable $x$, the corresponding hatted variable is defined as $\hat{x}=\frac{r_+ x}{l^2}$.}
\begin{align}
    &\left(\left(\cosh\hat{T}_{B}-\cosh(\hat{q}-\hat{t}_{0})\right)\partial_{\hat{q}}-\left(\cosh\hat{T}_{B}-\cosh(\hat{p}-\hat{t}_{0})\right)\partial_{\hat{p}}\right)g^R(\hat{q},\hat{p}) \nonumber \\ &=(h-1)\left(\sinh(\hat{p}-\hat{t}_{0})-\sinh(\hat{q}-\hat{t}_{0})\right) g^R(\hat{q},\hat{p})
\end{align}
and 
\begin{align}
    &\left(\left(\cosh\hat{T}_{B}-\cosh(\hat{p}^{new}-\hat{t}_{0})\right)\partial_{\hat{p}^{new}}-\left(\cosh\hat{T}_{B}-\cosh(\hat{q}^{new}-\hat{t}_{0})\right)\partial_{\hat{q}^{new}}\right)g^L(\hat{q}^{new},\hat{p}^{new}) \nonumber \\ &=(h-1)\left(\sinh(\hat{q}^{new}-\hat{t}_{0})-\sinh(\hat{p}^{new}-\hat{t}_{0})\right)g^L(\hat{q}^{new},\hat{p}^{new})\,.
\end{align}
Once again, we proceed by solving these equations using the method of characteristics. To obtain the corresponding results for the $A$ region, we set \(t_0=0\) and \(T_B=T_A\) and demand \([\tilde{H}_A^{(T)},\Phi(P)]=0\). After making these substitutions in the above equations and following an analysis similar to \cite{Kabat:2017mun} and in appendix \ref{app:outsidebtztechnical}, the most general solution is found to be\footnote{In writing the equations below, we have redefined ${q}-\frac{i\pi l^2}{r_+}\to {q}$ (and same with $p$-variables), where ${q}-\frac{i\pi l^2}{r_+}$ came naturally due to dealing with \([\tilde{H}_A^{(T)},\Phi(P)]=0\), and for taking the right boundary variables ${q}$ back to the left boundary by an imaginary time translation. We also had to redefine $q'={q}^{new}+\frac{i\pi l^2}{r_+}\to {q}^{new}$ (and same for $p$-variables).}
\begin{equation}
    g^R(\hat{q},\hat{p})=a_0 f_1(x)U^{h-1}\,\qquad\text{and}\qquad   g^L(\hat{q}^{new},\hat{p}^{new})=b_0 f_2(y)V^{h-1}\,,
\end{equation}
where 
\begin{equation}
  x=\frac{\sinh(\frac{\hat{T}_A+\hat{q}}{2})\sinh(\frac{\hat{T}_A+\hat{p}}{2})}{\sinh(\frac{\hat{T}_A-\hat{q}}{2})\sinh(\frac{\hat{T}_A-\hat{p}}{2})} \;,\;  y=\frac{\sinh(\frac{\hat{T}_A+\hat{q}^{new}}{2})\sinh(\frac{\hat{T}_A+\hat{p}^{new}}{2})}{\sinh(\frac{\hat{T}_A-\hat{q}^{new}}{2})\sinh(\frac{\hat{T}_A-\hat{p}^{new}}{2})}\,,
\end{equation}
   and
\begin{equation}
    U=\sinh\left(\frac{\hat{T}_A+\hat{q}}{2}\right)
    \sinh\left(\frac{\hat{T}_A+\hat{p}}{2}\right)
    \sinh\left(\frac{\hat{T}_A-\hat{q}}{2}\right)
    \sinh\left(\frac{\hat{T}_A-\hat{p}}{2}\right)\,,
\end{equation} 
\begin{equation}
	V=\sinh\left(\frac{\hat{T}_A+\hat{q}^{new}}{2}\right)\sinh\left(\frac{\hat{T}_A+\hat{p}^{new}}{2}\right)\sinh\left(\frac{\hat{T}_A-\hat{q}^{new}}{2}\right)\sinh\left(\frac{\hat{T}_A-\hat{p}^{new}}{2}\right)\,.
\end{equation}
 
However, we also impose
 \([\tilde{H}_{B}^{(T)},\Phi(P)]=0\) with the solutions obtained above to find out the functions \(f_{1}(x)\) and \(f_{2}(y)\). After some algebra, one can see that \(f_{1}(x)\) satisfies the following equation 
 \begin{equation}
    \frac{df_{1}}{dx}=\frac{h-1}{x}\frac{x-\alpha}{x+\alpha}f_{1}\,,
\end{equation}
and an equivalent equation for \(f_{2}(y)\) becomes
 \begin{equation}
    \frac{df_{2}}{dy}=\frac{h-1}{y}\frac{y-\alpha}{y+\alpha}f_{2}\,,
\end{equation}
with
\begin{equation}
\alpha=\frac{\sinh{\hat{t}_{0}}\sinh{\hat{T}_A}+\cosh{\hat{t}_{0}}\cosh{\hat{T}_A}-\cosh{\hat{T}_B}}{\sinh{\hat{t}_{0}}\sinh{\hat{T}_A}-\cosh{\hat{t}_{0}}\cosh{\hat{T}_A}+\cosh{\hat{T}_B}}\,.
\end{equation}
The above equations have solutions as
\begin{equation}
   f_1(x)=c_{1}\left(\frac{(x+\alpha)^2}{x}\right)^{h-1}
\end{equation}
and 
\begin{equation}
   f_2(y)=c_{2}\left(\frac{(y+\alpha)^2}{y}\right)^{h-1}\,.
\end{equation}
In order to obtain the solution in the required form, one can define a new variable \(\hat{t}_*\) which is nothing but the time at point \(P \) (in other words, defined using hindsight and obtained by solving for the intersection point of two spacelike parts of the extremal surfaces stretching between \((t_0-T_B,t_0+T_B)\) and \((-T_A,T_A)\)  beyond the horizon of the BTZ black holes \eqref{eq:btzext})
\begin{equation}
    \tanh\hat{t}_{*}=\frac{1}{\cosh\hat{t}_{0}}\left(\sinh\hat{t}_{0}-\frac{\cosh\hat{T}_{B}}{\sinh\hat{T}_{A}}\right)\,.
\end{equation}
In terms of \(\hat{t}_*\), \(\alpha\), \(g^{R}(\hat{q},\hat{p})\) and \(g^{L}(\hat{q}^{new},\hat{p}^{new})\) take the following values
\begin{equation}
\alpha=\frac{\sinh\hat{t}_*\sinh\hat{T}_B+\cosh\hat{t}_*\cosh\hat{T}_B}{\sinh\hat{t}_*\sinh\hat{T}_B-\cosh\hat{t}_*\cosh\hat{T}_B}\,,
\end{equation}
\begin{align}
    g^{R}(\hat{q},\hat{p})&=c_{1}\left(\cosh\left(\frac{\hat{p}-\hat{q}}{2}\right)+\frac{\cosh\hat{T}_{A}}{\sinh\hat{t}_{*}}\sinh\left(\frac{\hat{p}+\hat{q}}{2}-\hat{t}_{*}\right)\right)^{\Delta-2}\nonumber\\
    &=c_{1}\left(\cos{y}+\frac{\cosh\hat{T}_{A}}{\sinh\hat{t}_{*}}\sinh{x}\right)^{\Delta-2} \label{right}\,,
\end{align}
and
\begin{align}\label{eq:gL}
g^L(\hat{q}^{new},\hat{p}^{new})&=c_{2}\left(\cosh\left(\frac{\hat{p}^{new}-\hat{q}^{new}}{2}\right)-\frac{\cosh\hat{T}_{A}}{\sinh\hat{t}_{*}}\sinh\left(\frac{\hat{p}^{new}+\hat{q}^{new}}{2}+\hat{t}_{*}\right) \right)^{\Delta-2}\nonumber\\
&=c_{2}\left(\cos{y}-\frac{\cosh\hat{T}_{A}}{\sinh\hat{t}_{*}}\sinh{x}\right)^{\Delta-2}\,.
\end{align}

The region of integration is determined in the same way as the exterior case discussed in the last section. Thus for the right boundary smearing function \( g^R(\hat{q},\hat{p})\), the region of integration is defined as
\begin{equation}\label{eq:rbdryconst}
    \cos{y}+\frac{\cosh\hat{T}_{A}}{\sinh\hat{t}_{*}}\sinh{x}>0\,.
\end{equation}
And for the left boundary, with a boundary support given by 
\begin{equation}\label{eq:lbdryconst}
    \cos{y}-\frac{\cosh\hat{T}_{A}}{\sinh\hat{t}_{*}}\sinh{x}<0\,.
\end{equation}
In the above equation \eqref{eq:gL}, we can go back to the coordinates \((q',p')\) by performing an inverse transformation to what we have done before. This will readily give us the HKLL prescription for sub-horizon bulk fields \eqref{27in}. In fact, in writing \eqref{eq:lbdryconst} as well, we have converted an inequality like \eqref{eq:rbdryconst} back to the left boundary.

\section{de Sitter reconstruction\label{sect:dS}}

We now turn to the issue of bulk reconstruction in de Sitter (dS) spacetimes using the techniques of modular Hamiltonians. As we mentioned in the introduction, the timelike entanglement in AdS/CFT were originally motivated from the studies of pseudo-entropy in dS, which turns out to be the natural, complex, entropy associated to a Euclidean boundary subregion. So, it is not a surprise that pseudo-entropies can be a natural candidate when it comes to bulk reconstruction using the entanglement structure of the spacetime. We provide a support to this statement by considering bulk reconstruction in dS flat slicings. The HKLL reconstruction for some of these patches were already carried out in \cite{Xiao:2014uea} for scalars, and for higher spins in \cite{Sarkar:2014dma}. Our primary aim here will be to re-derive these results, but this time starting from the modular Hamiltonian corresponding to pseudo-entropy. Our computation here is not completely airtight, in that there are some assumptions, for which a bulk insight is definitely needed. But we can nonetheless have a boundary reconstruction in various slicings of dS. The computation that appears in this section can also be carried out in the static patch of dS, but we have excluded them here as they can be carried out in a very similar manner.

\subsection{Flat slicing\label{subsec:dSflat}}

\subsubsection{Invariance under timelike modular flow\label{subsec:dSflatinv}}

We start with the flat slicing of dS, obtained by analytically continuing the spacetime from AdS Poincar\'{e} Patch \eqref{201}. One requires to perform a double Wick rotation
\begin{equation}\label{eq:dSAdSancont}
    z \to -i\eta \;,\; t \to -it_E\;,\;R_{AdS}\to -il_{dS}\,,
\end{equation} 
under which the metric becomes 
\begin{equation}
    ds^2=\frac{l^2_{dS}}{\eta^2}(-d\eta^2+dx^2+dt_E^2)\,.
\end{equation} 
In other words, the radial coordinate now becomes the emergent time direction $\eta$, whereas the boundary coordinates are consisted of $(x,t_E)$ and it is now of Euclidean signature. As was emphasized in \cite{Xiao:2014uea}, from the point of bulk microcausality, in dS we require to consider both normalizable \(\mathcal{O}_+\) and non-normalizable \(\mathcal{O}_-\) modes of a bulk field, which in this case are simply the positive and negative frequency modes. In particular, we have 
\begin{eqnarray}
   \Phi(t_E,x,\eta)=\mathcal{A}\int_{\sigma>0}dx'\,dt'_E\,\left(\frac{\eta^2-x'^2-t_E'^2}{\eta}\right)^{\Delta-2}\mathcal{O}_+(x+x',t_E+t_E')\nonumber\\
    + \mathcal{B}\int_{\sigma>0}dx'\,dt'_E\,\left(\frac{\eta^2-x'^2-t_E'^2}{\eta}\right)^{-\Delta}\mathcal{O}_-(x+x',t_E+t_E')
\end{eqnarray}  
where 
\begin{equation}
    \mathcal{A}=\frac{\Gamma(\Delta)}{\pi \Gamma(\Delta-1)}\;,\;
\mathcal{B}=\frac{\Gamma(2-\Delta)}{\pi \Gamma(1-\Delta)}\,,
\end{equation}
and $\sigma>0$ simply denotes the causal support of the bulk field, in retarded (or advanced) time. Here $\sigma$ is the bulk to bulk dS covariant distance given by 
\begin{equation}
    \sigma=\frac{{\eta}^{2}+{\eta'}^{2}-({x}^{2}-{x'}^{2})-(t_{E}^{2}-t'^2_{E})}{2\eta \eta'}\,.
\end{equation}
Introducing Euclidean lightcone coordinates \(\xi=t_E -ix'\) and \(\bar{\xi}=t_E+ix'\) and taking \(x=0\) plane, we get ($\mathcal{A}$, $\mathcal{B}$ are some spacetime and conformal dimension dependent constants)
\begin{align}\label{eq:dsscalarmod}
    \Phi(t_E,x,\eta)&= \mathcal{A}\int_{\sigma>0}dx'\,dt'_E\,\left(\frac{\eta^2-x'^2-t_E'^2}{\eta}\right)^{\Delta-2}e^{t'_E\partial_{t_E}}\mathcal{O}_+(x',t_E)\nonumber\\ &+ \mathcal{B}\int_{\sigma>0}dx'\,dt'_E\,\left(\frac{\eta^2-x'^2-t_E'^2}{\eta}\right)^{-\Delta}e^{t'_E\partial_{t_E}}\mathcal{O}_-(x',t_E)\nonumber\\
    &=\Phi_++\Phi_-\,.
\end{align}
The next step is to write down the corresponding modular Hamiltonian of a subregion at  \(x=0\) slice and its endpoints defined by 
\((t_{1E},t_{2E})\) \cite{Cardy:2016fqc}.\footnote{Of course, there is no a priori distinction between $x$ and $t_E$. However, our choice can be understood as defining which directions we are calling $x$ and which one $t_E$. Also, as done previously, we have extended the limits of the $\xi$ integrals in the Euclidean timelike modular Hamiltonians, to give it an `extended' look. It would once again be interesting to investigate its possible algebraic origin.} We have
\begin{equation}
    \tilde{H}_{dS}=2\pi \int_{-\infty}^{\infty} \frac{(t_{2E}-\xi)(\xi - t_{1E})}{(t_{2E}-t_{1E})}T_{\xi \xi}(\xi) \,d\xi + \text{anti-chiral part}\,.
\end{equation} 
One other way to obtain this could be to start from \eqref{eq:LSLmodham}, which is the spacelike modular Hamiltonian in Lorentzian CFT (dual to AdS spacetimes), and then consider the analytic continuation given in \eqref{eq:dSAdSancont} (along with the proper definition of lightcone coordinates and taking the $L\to\infty$ limit).\footnote{One might also think of starting from \eqref{eq:slmh} or \eqref{eq:tlmhbtz}, which are modular Hamiltonians at finite temperatures, and take $\beta\to\infty$ limit.}
Because in this section, we will only be talking about these modular Hamiltonians (these are not really `timelike', as they have been defined on Euclidean CFTs. Maybe a better name for them is pseudo-modular Hamiltonian or Euclidean modular Hamiltonian), we have skipped the usual superscript $(T)$ on the $\tilde{H}$ (like we had for AdS) to avoid clutter. For a primary operator of conformal weight \(\Delta\), which is dual to the bulk field $\Phi$, the dS/CFT dictionary suggests that $\Delta$ is also the dimension of $\mathcal{O}_+$. We can then derive
\begin{equation}
    [\tilde{H}_{dS}, \mathcal{O}_+(\xi,\bar{\xi})] = \frac{2\pi}{(t_{2E}-t_{1E})}\left(\Delta(\bar{\xi}-\xi)-t_{1E}t_{2E}(\partial_{\xi}-\partial_{\bar{\xi}})+(t_{1E}+t_{2E})(\xi \partial_{\xi}-\bar{\xi}\partial_{\bar{\xi}})+\bar{\xi}^2\partial_{\bar{\xi}}-\xi^2 \partial_{\xi} \right)\mathcal{O}_+\,.
\end{equation}
On the other hand, the operator $\mathcal{O}_-$ has the shadow dimension $(2-\Delta)$ for dS$_3$/CFT$_2$. Using that, we have
\begin{align}
&[\tilde{H}_{dS},\mathcal{O}_{-}(\xi,\bar{\xi})]\nonumber\\ 
 & = \frac{2\pi}{(t_{2E}-t_{1E})}\left((2-\Delta)(\bar{\xi}-\xi)-t_{1E}t_{2E}(\partial_{\xi}-\partial_{\bar{\xi}})+(t_{1E}+t_{2E})(\xi \partial_{\xi}-\bar{\xi}\partial_{\bar{\xi}})+\bar{\xi}^2\partial_{\bar{\xi}}-\xi^2 \partial_{\xi} \right)\mathcal{O}_{-}\,.
\end{align}
As in the AdS case, we can now introduce new coordinates \(q=\xi+t'_E\) and \(p=\bar{\xi}+t'_E\), in terms of which we have
\begin{align}
[\tilde{H}_{dS},\Phi]&=\frac{2\pi }{(t_{2E}-t_{1E})}\biggl(\mathcal{A}\int \left(\frac{\eta^2-(q-t_E)(p-t_E)}{\eta}\right)^{\Delta-2} \biggl(\Delta(p-q)-t_{1E}t_{2E}(\partial_{q}-\partial_{p})\nonumber \\ &+(t_{1E}+t_{2E})(q \partial_{q}-p\partial_{p})+p^2\partial_{p}-q^2\partial_{q}\biggl)\mathcal{O}_{+}+\mathcal{B}\int \left(\frac{\eta^2-(q-t_E)(p-t_E)}{\eta}\right)^{-\Delta}\nonumber \\ &\biggl((2-\Delta)(p-q)-t_{1E}t_{2E}(\partial_{q}-\partial_{p})+(t_{1E}+t_{2E})(q \partial_{q}-p\partial_{p})+p^2\partial_{p}-q^2\partial_{q}\biggl)\mathcal{O}_{-}\biggr)\,.
\end{align}

\begin{figure}[h]
    \centering
    \includegraphics[scale=0.5]{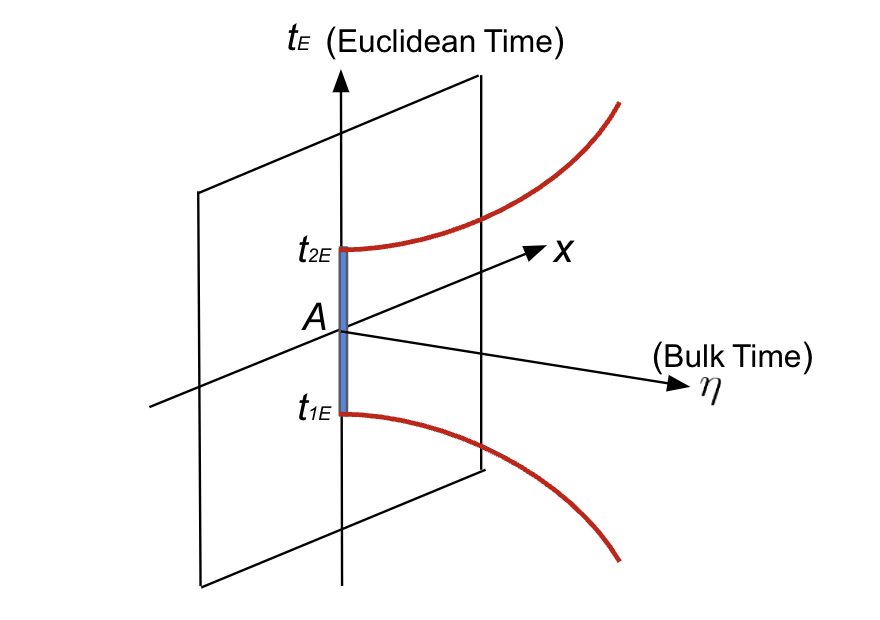}
    \caption{Timelike extremal surfaces in dS$_3$/CFT$_2$ for a subregion located at $x=0$ and endpoints at $t_{1E}$ and $t_{2E}$.}
    \label{fig:ds_01}
\end{figure}

We will now see that if we once again impose the boundary constraint that the bulk field commutes with the corresponding pseudo-modular Hamiltonians, one can recover that the dS bulk fields, as given in \eqref{eq:dsscalarmod}, must be located on the bulk geodesics emanating from the subregion described above (see figure \ref{fig:ds_01}). Indeed we obtain 
\begin{align}
&[\tilde{H}_{dS},\Phi]\nonumber\\
&=  \frac{2\pi\left(\eta^2-t_{1E}t_{2E}+(t_{1E}+t_{2E})t_{E}-{t_{E}}^2 \right) }{(t_{2E}-t_{1E})}\biggl(\mathcal{A}(\Delta-2)\int\left(\frac{{\eta}^2-(q-t_{E})(p-t_E)}{\eta}\right)^{\Delta-3}\left(\frac{p-q}{\eta}\right)\mathcal{O}_{+}(q,p) \nonumber \\ &-\mathcal{B}\Delta\int\left(\frac{\eta^2-(q-t_E)(p-t_E)}{\eta}\right)^{-\Delta-1}\left(\frac{p-q}{\eta}\right)\mathcal{O}_{-}(q,p)\biggr) =0\,.
\end{align}
It is clear that if
\begin{equation}
    \eta^2-t_{1E}t_{2E}+(t_{1E}+t_{2E})t_E-t_E^2=0\,,
\end{equation}
then the above equation is satisfied.
This turns out to be the timelike geodesic associated with a spacelike subregion in a Euclidean CFT dual to dS space \cite{Doi:2022iyj}. Given a subregion with endpoints \(t_{1E}={-T_0}\) and \(t_{2E}=+{T_0}\), the geodesic equation condenses to the form
\begin{equation}
    \eta^2+{T_0^2}-t_E^2=0\,.
\end{equation}

\subsubsection{Re-deriving HKLL in dS\label{subsec:dSflathkll}}

\begin{figure}[h]
    \centering
    \includegraphics[scale=0.6]{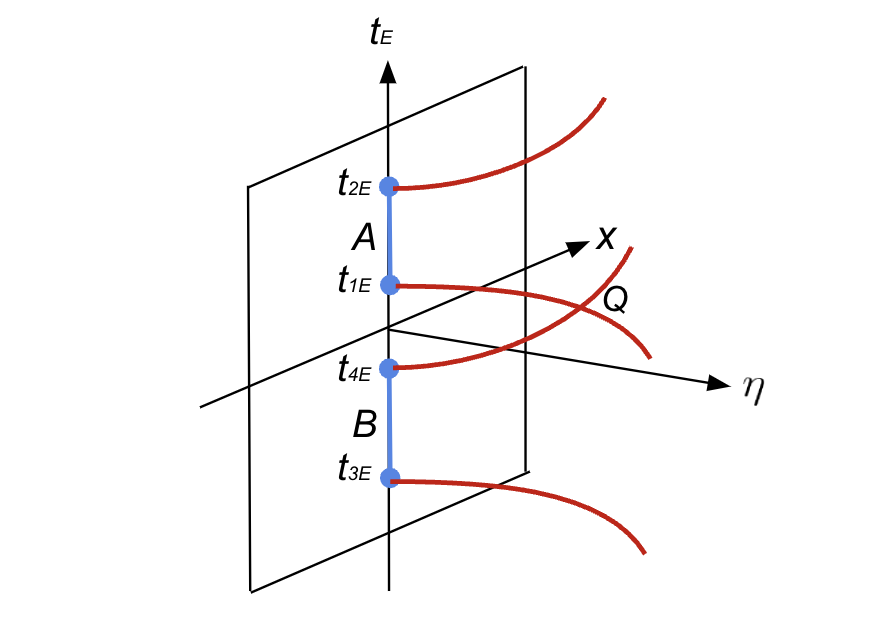}
    \caption{Two timelike geodesics intersect each other at a point $Q$ in the bulk, where the bulk field will be located.}
    \label{fig:ds_02}
\end{figure}

We now turn to use the invariance of timelike geodesic under the flow of Euclidean modular Hamiltonian, and set out for \emph{deriving} the HKLL bulk fields in dS. We will closely follow the steps that we performed for its \(AdS\) counterpart. We start by placing this bulk field at the intersection point of two \emph{non-overlapping} spacelike subregions, so that the corresponding timelike geodesics intersect each other at a point $Q$ in the bulk. See figure \ref{fig:ds_02}.

Making the ansatz\footnote{As we mentioned in the beginning of this section, purely from a boundary perspective, we currently lack an independent argument behind such a proposal. Any argument which hints that we require two sets of fields $\mathcal{O}_{\pm}$, hinges on some bulk insights one way or the other.}
\begin{equation}
    \Phi(t_E,\eta,x=0) = \int dx'dt'_E\,g_1(q,p)\,\mathcal{O}_+(q,p)+\int dx'dt'_E\,g_2(q,p)\,\mathcal{O}_-(q,p)
\end{equation}
We start off by imposing 
\begin{equation}
    (t_{1E}-t_{2E})[\tilde{H}_{A,dS},\Phi]=(t_{3E}-t_{4E})[\tilde{H}_{B,dS},\Phi]=0\,,
\end{equation}
where the subscripts $A$ and $B$ denotes the names of the subregions. Evaluating the subtraction between the two commutators above, we get
\[\int dq dp\,g_1(q,p)\biggl((t_{3E}t_{4E}-t_{1E}t_{2E})(\partial_{q}-\partial_{p})+(t_{1E}+t_{2E}-t_{3E}-t_{4E})(q\partial_{q}-p\partial_{p})\biggr)\mathcal{O}_+\]
\begin{equation}
    +\int dq dp\,g_2(q,p)\biggl((t_{3E}t_{4E}-t_{1E}t_{2E})(\partial_{q}-\partial_{p})+(t_{1E}+t_{2E}-t_{3E}-t_{4E})(q\partial_{q}-p\partial_{p}) \biggr)\mathcal{O}_-=0\,.
\end{equation}
We proceed again by doing integration by parts to arrive at 
\[ \int \mathcal{O}_+ \biggl((t_{3E}t_{4E}-t_{1E}t_{2E})(\partial_{q}g_1-\partial_{p}g_1)+(t_{1E}+t_{2E}-t_{3E}-t_{4E})(g_1\partial_{q}-g_1\partial_{p}+q\partial_{q}g_1+p\partial_{p}g_1)\biggr)\]
\begin{equation}
    +\int \mathcal{O}_- \biggl((t_{3E}t_{4E}-t_{1E}t_{2E})(\partial_{q}g_2-\partial_{p}g_2)+(t_{1E}+t_{2E}-t_{3E}-t_{4E})(g_2\partial_{q}-g_2\partial_{p}+q\partial_{q}g_2+p\partial_{p}g_2)\biggr)=0\,.
\end{equation} 
These are two independent integrals, so the only way the whole term is zero is if they are individually equal to zero. Then the two solutions are 
\begin{equation}
    g_1(q,p)=f_1((q-T_E)(p-T_E))\;,\;g_2(q,p)=f_2((q-T_E)(p-T_E)) 
\end{equation} 
where \(T_E=\frac{t_{1E}t_{2E}-t_{3E}t_{4E}}{t_{1E}+t_{2E}-t_{3E}-t_{4E}}\).
Solving them by a similar method of characteristics just like in AdS, we finally get 
\begin{equation}
    g_1(q,p)=c_{\Delta}\left(\eta^2-(q-T_E)(p-T_E)\right)^{\Delta-2}
\end{equation} 
and 
\begin{equation}
    g_2(q,p)=c_{2-\Delta}\left(\eta^2-(q-T_E)(p-T_E)\right)^{-\Delta}
\end{equation}
along with 
\begin{equation}
    \eta^2-t_{1E}t_{2E}+(t_{1E}+t_{2E})T_E-T_E^2=0\,.
\end{equation}
The free parameters \(\eta\) and \(T_E\) define the coordinates of the local bulk field with $\eta$ providing the emergent time direction. 

An exact similar analysis (more like what we did for outside horizon case of BTZ black holes) can also be carried out for dS static patches. And unsurprisingly, we can draw the same conclusions as above for that case as well. We will not include those calculations here as they can be carried out in the same manner.

\section{Conclusion and outlook}\label{sec:concl}

The main message of this paper was to show that the entanglement of the boundary CFT, especially in two-dimensions, can be used to capture the physics of semiclassical free bulk fields in the dual AdS background. In particular, we have reconstructed, in examples of BTZ black holes, the physics \emph{behind} the black hole horizons upon knowing only a part of the CFT state and its associated entanglement structure. The timelike entanglement seems to play a necessary and crucial role in this, which also elucidates how one can go about reconstructing local bulk physics in de Sitter. Our work therefore has some relevance in black hole information problem in AdS (and by extension, hopefully, for asymptotically flat ones), and for cosmological horizons.

There are some immediate aspects that one may wish to discuss in light of timelike entanglement. In particular, we relied heavily on the symmetry of CFT$_2$ in order to find out the \emph{bulk} timelike modular Hamiltonian. This gave us confidence to impose the \emph{boundary} constraint that bulk fields at certain locations (i.e. on the spacelike part of the extremal surface homologous to the timelike boundary) commute with the timelike modular Hamiltonian at the boundary. While we expect it to be true also in higher dimensions, it is definitely more non-trivial to argue for such a property explicitly and purely from the higher dimensional boundary CFT's perspective. In general, even for spacelike entanglement, the analysis is technically more challenging for higher dimensions, as it requires more (than two) numbers of intersecting RT surfaces to localize a field in the bulk. We have also assumed states with a translational symmetry along the boundary spatial directions, and it will be interesting to try and extend these methods for more general states.

Another relevant question is the study of modular flow at the red lined parts of the extremal surface (i.e. the timelike part). These surfaces are non-invariant under the timelike modular flow. However, upon knowing how exactly should one expect the bulk fields to transform, it should be in principle possible to reconstruct a bulk field at a point where the green and red line intersects (e.g.~in figures \ref{fig:BTZ_02} and \ref{fig:BTZ_05}). It will also be interesting to see if one can construct the modular Hamiltonian for a union of subregions (e.g.~in the left and right CFTs, much like in \cite{Hartman:2013qma}, and for an analogous case with timelike subregions), and use that to evaluate bulk fields in the interior of the horizon. We leave these questions for future investigations.

\bigskip
\goodbreak
\centerline{\bf Acknowledgements}
\noindent
We are grateful to Daniel Kabat for his valuable inputs towards the draft. DS would also like to thank the hospitality and lively discussions that were provided by IAS, Tshinghua University during the final stages of this work. The work of DS is supported by the DST-FIST grant number SR/FST/PSI-225/2016 and SERB MATRICS grant MTR/2021/000168.

\appendix

\section{Modular flow due to timelike modular Hamiltonian}\label{app:app}

This appendix is devoted to an explicit construction of the boundary and bulk timelike modular Hamiltonians, utilizing the conformal symmetry generators of CFT$_2$. This helps us understand the flow of any bulk field under these modular Hamiltonians which we have used in the main part of the paper for the bulk reconstruction of scalar fields.

\subsection{Modular flow for CFT$_2$ in flat space and AdS$_3$ Poincar\'{e}}\label{appsec:modflowpureads}

The construction of modular Hamiltonian for a subregion (either timelike or spacelike) stretched between points $x^\mu = (a , \bar{a})$  and $y^\mu = (b , \bar{b})$ in the lightcone coordinates of a CFT in flat spacetime can be written very generally. In what follows, the CFT is located in the conformal boundary of the AdS$_3$ Poincar\'{e} patch \eqref{201}
\begin{equation}
    ds^2 = \frac{l^2}{z^2} (-dt^2 + dx^2 + dz^2) = \frac{l^2}{z^2}(dz^2 -d\omega d\bar{\omega})\label{@m14}
\end{equation}
with the boundary lightcone coordinates defined as
\begin{equation}\label{eq:lcother}
	\omega = t +x\,,\hspace{1cm} \bar{\omega}= t-x\,.
\end{equation}

\subsubsection{Boundary flow}

In this case, the boundary (extended) modular Hamiltonian and modular momentum take the form \cite{deBoer:2016pqk,Kabat:2017mun,Czech:2019vih,Banerjee:2022jnv}
\begin{equation}
    \tilde{H}_{mod}= K_+ + K_- \qquad\text{and}\qquad \tilde{P}_{mod}= K_+ - K_- \label{@m1}\,,
\end{equation}
where
\begin{equation}
    K_+ = s_1L_1 + s_0L_0 + s_{-1}L_{-1}\qquad\text{and}\qquad  K_- = t_1\bar{L}_1 + t_0\bar{L}_0 + t_{-1}\bar{L}_{-1}\label{@m3}\,.
\end{equation}
The $s_i$ and $t_i$'s are defined in terms of the subregion size as 
\begin{equation}
    s_1 = \frac{2\pi i }{(b-a)}\,, \hspace{1cm}  t_1 = - \frac{2\pi i}{(\bar{b}-\bar{a})}\,,\label{@m6}
\end{equation}
\begin{equation}
    s_0 = -{2\pi i }\Bigg(\frac{b+a}{b-a}\Bigg)\,,\hspace{1cm}  t_0 ={2\pi i }\Bigg(\frac{\bar{b}+\bar{a}}{\bar{b}-\bar{a}}\Bigg)\,,\label{@m7}
\end{equation}
\begin{equation}
    s_{-1} = \frac{2\pi i (a.b) }{(b-a)}\,,\hspace{1cm}  t_{-1} = -  \frac{2\pi i (\bar{a}.\bar{b}) }{(\bar{b}-\bar{a})}\,.\label{@m8}
\end{equation}
Whereas in the Poincar\'{e} patch, the representations of Virasoro generators are usually taken to be of the form 
\begin{equation}
    L_{-1} = i \partial_\omega, \hspace{1cm} L_0 = -\omega \partial_\omega \hspace{0.5cm}\text{and} \hspace{0.5cm} L_{1} = - i\omega^2\partial_\omega\,,
    \label{@m4}
\end{equation}
with
\begin{equation}
    \bar{L}_{-1} = i \partial_{\bar{\omega}} , \hspace{1cm} \bar{L}_0 = -\bar{\omega} \partial_{\bar{\omega}}\hspace{0.5cm}\text{and} \hspace{0.5cm} \bar{L}_{1} = - i \bar{\omega}^2\partial_{\bar{\omega}}\,. \label{@m5}
\end{equation}

\subsubsection{Bulk flow}

Written in this way, it is now immediate to uplift the (extended) modular Hamiltonians and modular momentums to their corresponding bulk counterparts. They take the form \cite{Balasubramanian:1998sn,Keski-Vakkuri:1998gmz,Das:2019iit,Das:2020goe}
\begin{equation}
    \tilde{H}_{mod}^{bulk}= K_+^{bulk} + K_-^{bulk} \qquad\text{and}\qquad\tilde{P}_{mod}^{bulk}= K_+^{bulk} - K_-^{bulk}\,,\label{@m9}
\end{equation}
where
\begin{equation}
    K_+^{bulk} = s_1L_{b,1} + s_0L_{b,0} + s_{-1}L_{b, -1}\qquad\text{and}\qquad K_-^{bulk} = t_1\bar{L}_{b,1} + t_0\bar{L}_{b,0} + t_{-1}\bar{L}_{b,-1}\,.\label{@m11}
\end{equation}
$L_{b,n}$'s are the bulk AdS$_3$ isometry generators corresponding to the boundary Virasoro generators $L_n$'s. They take the form\footnote{The subscript $b$ in $L_b$ above will always denote `bulk', and it is not to be confused with the endpoint coordinates $b,\bar{b}$.}
\begin{equation}
    L_{b,-1} = i \partial_\omega, \hspace{1cm} L_{b,0} = - \frac{z}{2} \partial_z - \omega\partial_{\omega} \hspace{0.5cm}\text{and} \hspace{0.5cm} L_{b,1} = - i\bigg(\omega z \partial_z + \omega^2 \partial_{\omega} + z^2 \partial_{\bar{\omega}}\bigg)\,,\label{@m12}
\end{equation}
with
\begin{equation}
    \bar{L}_{b,-1} = i \partial_{\bar{\omega}} , \hspace{1cm} \bar{L}_{b,0} =  - \frac{z}{2} \partial_z -\bar{\omega}\partial_{\bar{\omega}}\hspace{0.5cm}\text{and} \hspace{0.5cm} \bar{L}_{b,1} = - i\bigg(\bar{\omega} z \partial_z + \bar{\omega}^2 \partial_{\bar{\omega}} + z^2 \partial_{\omega}\bigg)\,.\label{@m13}
\end{equation}

\subsubsection{Intervals on spatial and temporal slices}

Given these general formulas it is now straightforward to write down the boundary and bulk extended modular Hamiltonians for subregions of various types. 
\vspace{1mm}

\noindent
\emph{(i) Spacelike intervals:} For a purely spacelike interval stretched between two endpoints $P ( x = -R, t =0 )$ and $Q ( x = +R , t=0)$, using the above formulas we immediately obtain the boundary operators 
\begin{equation}
    \tilde{H}_{mod}:= \frac{\pi i}{R} \left[(L_1 + \bar{L}_1) -  R^2 ( L_{-1} + \bar{L}_{-1})\right]=\frac{\pi}{R}\left[\big(R^2 -x^2 - t^2\big)\,\partial_t - 2xt\,\partial_x\right]\label{@a}
\end{equation}
and
\begin{equation}
    \tilde{P}_{mod}:= \frac{\pi i}{R} \left[(L_1 - \bar{L}_1) -  R^2 ( L_{-1} -\bar{L}_{-1})\right]=\frac{\pi}{R}\left[-\big(R^2 -x^2 - t^2\big)\partial_x + 2xt\,\partial_t\right]\,.\label{@b}
\end{equation}
Similarly the corresponding bulk operators are
\begin{equation}
    \tilde{H}_{mod}^{bulk}:= \frac{\pi i}{R} \bigg[(L_{b,1} + \bar{L}_{b,1}) -  R^2 ( L_{b,-1} + \bar{L}_{b,-1})\bigg]= \frac{\pi}{R}\left[\big(R^2 -x^2 - t^2\big)\partial_t - 2xt\,\partial_x - z^2\partial_t - 2zt\,\partial_z\right] \label{@c}
\end{equation}
and
\begin{equation}
    \tilde{P}_{mod}^{bulk}:= \frac{\pi i}{R} \bigg[(L_{b,1} - \bar{L}_{b,1}) -  R^2 ( L_{b,-1} - \bar{L}_{b,-1})\bigg]=\frac{\pi}{R}\left[-\big(R^2 -x^2 - t^2\big)\partial_x + 2xt\,\partial_t - z^2\partial_x - 2zx\,\partial_z\right]\,.
\end{equation}
These results are of course known, and they clearly indicate that for $t=0$, $x =\pm R$ points remain invariant under the boundary modular flow (using \eqref{@a} and \eqref{@c}). While the bulk RT surface $\sqrt{x^2 + z^2 } = R$ remains invariant under bulk modular flow. One can also check that the action of $\tilde{H}_{mod}$ and $\tilde{P}_{mod}$ are orthogonal to each other by computing $[\tilde{H}_{mod},\tilde{P}_{mod}] = 0$ (using \eqref{@a} and \eqref{@b}).
\vspace{1mm}

\noindent
\emph{(ii) Timelike intervals:} On the other hand, for a purely timelike interval with two endpoints $P (t = -T_0, x = 0)$ and $Q ( t= +T_0, x = 0 )$, the resulting boundary modular Hamiltonian and modular momentum are given by 
\begin{equation}
    \tilde{H}_{mod}^{(T)}:= \frac{\pi i}{T_0} \bigg[(L_1 - \bar{L}_1) -  T_0^2 ( L_{-1} - \bar{L}_{-1})\bigg]=\frac{\pi}{T_0}\left[\big(T_0^2 -x^2 - t^2\big)\partial_x - 2xt\,\partial_t\right]\label{@d}
\end{equation}
and
\begin{equation}
    \tilde{P}_{mod}^{(T)}:= \frac{\pi i}{T_0} \bigg[(L_1 + \bar{L}_1) -  T_0^2 ( L_{-1} + \bar{L}_{-1})\bigg]=\frac{\pi}{T_0}\left[-\big(T_0^2 -x^2 - t^2\big)\partial_t + 2xt\,\partial_x\right]\,.
\end{equation}
We clearly see that the endpoints of the timelike intervals are invariant under this Hamiltonian. Their counterparts in the bulk then takes the form
\begin{equation}
    \tilde{H}_{mod}^{bulk (T)}:= \frac{\pi i}{T_0} \bigg[(L_{b,1} - \bar{L}_{b,1}) -  T_0^2 ( L_{b,-1} - \bar{L}_{b,-1})\bigg]=\frac{\pi}{T_0}\left[\big(T_0^2 -x^2 - t^2\big)\partial_x - 2xt\,\partial_t + z^2\partial_x + 2zx\,\partial_z\right]\label{@e}
\end{equation}
and
\begin{equation}
    \tilde{P}_{mod}^{bulk (T)}:= \frac{\pi i}{T_0} \bigg[(L_{b,1} + \bar{L}_{b,1}) -  T_0^2 ( L_{b,-1} + \bar{L}_{b,-1})\bigg]=\frac{\pi}{T_0}\left[-\big(T_0^2 -x^2 - t^2\big)\partial_t + 2xt\,\partial_x + z^2\partial_t + 2zt\,\partial_z\right]\,.
\end{equation}
This time, we see that the spacelike parts of the extremal surface for a timelike subregion (i.e.~the green lines in figure \ref{fig:pure})
 \begin{equation}\label{eq:slinv}
 	t=\sqrt{T_0^2 + z^2 }
 \end{equation}
remain invariant under the bulk modular flow, but not the timelike parts (i.e.~the red line). This Information will help us formulate the bulk reconstruction problem using timelike modular Hamiltonian in the main part of the paper.

\subsection{Modular flow for thermal CFT$_2$ and in BTZ}

\subsubsection{Boundary flow}

It is straightforward to carry out a similar computation for a thermal CFT. Here we will only describe the case for BTZ black holes using the metric given in \eqref{eq:btzmet}, from which the AdS$_3$ Rindler case can follow quite simply. In this case the resulting formulas are \cite{Keski-Vakkuri:1998gmz,Banerjee:2022jnv}
\begin{equation}
    \tilde{H}_{mod}= K_+ + K_-, \hspace{1cm} \tilde{P}_{mod} = K_+ - K_-\label{@m55}
\end{equation}
with 
\begin{equation}
    K_+ = s_1L_1 + s_0L_0 + s_{-1}L_{-1}\qquad\text{and} \qquad K_- = t_1\bar{L}_1 + t_0\bar{L}_0 + t_{-1}\bar{L}_{-1}\,.
\end{equation}
The standard representation to use for the global Virasoro generators are\footnote{Here we have defined dimensionless temporal and angular coordinates $\hat{t} = \frac{r_+ t}{l^2}$ and $\hat{\phi}=\frac{r_+ \phi}{l}$\,, with the boundary lightcones defined in terms of them. In particular, $\hat{x}^+ = \hat{t} + \hat{\phi}$ and $\hat{x}^- = \hat{t} - \hat{\phi}$.}
\begin{equation}
    L_{-1} = - e^{- \hat{x}^+}\partial_+ , \hspace{1cm} L_0 = - \partial_+ \hspace{0.5cm}\text{and} \hspace{0.5cm} L_{1} = - e^{\hat{x}^+}\partial_+\label{@m58}
\end{equation}
and
\begin{equation}
    \bar{L}_{-1} = - e^{-\hat{x}^-}\partial_- , \hspace{1cm} \bar{L}_0 = - \partial_- \hspace{0.5cm}\text{and} \hspace{0.5cm} \bar{L}_{1} = - e^{\hat{x}^-}\partial_-\label{@m59}\,.
\end{equation}
The $s,t$ coefficients are given by\footnote{Definition of $a^+$ and $a^-$ are $a^{\pm}=t\pm l\phi$. And similarly for $b$.}
\begin{equation}
    s_1 = \frac{2 \beta  \coth{\left(\frac{2\pi}{\beta}\frac{b^+ - a^+}{2}\right)}}{e^{\frac{2\pi}{\beta}a^+} + e^{\frac{2\pi}{\beta}b^+}}\,, \hspace{1cm}  t_1 = - \frac{2\beta \coth{\left(\frac{2\pi}{\beta}\frac{b^- - a^-}{2}\right)}}{e^{\frac{2\pi}{\beta}a^-} + e^{\frac{2\pi}{\beta}b^-}}\,,\label{@m60}
\end{equation}
\begin{equation}
    s_0 = -{2\beta \coth{\left(\frac{2\pi}{\beta}\frac{b^+ - a^+}{2}\right)}}\,,\hspace{1cm}  t_0 ={2\beta \coth{\left(\frac{2\pi}{\beta}\frac{b^- - a^-}{2}\right)}}\,,\label{@m61}
\end{equation}
\begin{equation}
    s_{-1} = \frac{2\beta \coth{\left(\frac{2\pi}{\beta}\frac{b^+ - a^+}{2}\right)}}{e^{-\frac{2\pi}{\beta}a^+} + e^{-\frac{2\pi}{\beta}b^+}}\,, \hspace{1cm}  t_{-1} = - \frac{2\beta \coth{\left(\frac{2\pi}{\beta}\frac{b^- - a^-}{2}\right)}}{e^{-\frac{2\pi}{\beta}a^-} + e^{-\frac{2\pi}{\beta}b^-}}\,.\label{@m62}
\end{equation}

Our next goal is to uplift the Virasoro generators to the bulk in order to write the corresponding extended bulk modular Hamiltonians. Because we are using the coordinates as in \eqref{eq:btzmet}, we will derive the expressions for the modular flow at points inside and outside the horizon separately.

Once again, we discuss the two types of intervals separately:
\vspace{1mm}

\noindent
\emph{(i) Spacelike intervals:} For a spatial subregion stretched between two points $P ({\phi} = -R/l, {t} =0 )$ and $Q ( {\phi} = +R/l , {t}=0)$, the resulting $s,t$ coefficients yield a boundary modular Hamiltonian (and momentum) given by 
\begin{equation}
    \tilde{H}_{mod}:= \frac{\beta}{\sinh{\frac{r_+ R}{l^2}}} \bigg[(L_1 +L_{-1} ) +  ( \bar{L}_1 +\bar{L}_{-1}) - 2 \cosh{\frac{r_+ R }{l^2}}(L_0 + \bar{L}_0)\bigg]\qquad\text{and}
\end{equation}
\begin{equation}
    \tilde{P}_{mod}:=\frac{\beta}{\sinh{\frac{r_+ R}{l^2}}} \bigg[(L_1 + L_{-1}) -  ( \bar{L}_1 +\bar{L}_{-1}) - 2 \cosh{\frac{r_+ R }{l^2}}(L_0 - \bar{L}_0)\bigg]\,.
\end{equation}
Using \eqref{@m58} and \eqref{@m59}, these take the following form
\begin{equation}
    \tilde{H}_{mod}:=\frac{\beta}{\sinh{\frac{r_+ R}{l^2}}}\left[\left( \cosh{\frac{r_+ R}{l^2}} - \cosh{\hat{t}}\cosh{\hat{\phi}}\right)\partial_{{t}} - \frac{1}{l}\sinh{\hat{t}}\,\sinh{\hat{\phi}} \,\partial_{{\phi}}\right]\label{@107}
\end{equation}
and 
\begin{equation}
    \tilde{P}_{mod}:= \frac{\beta}{\sinh{\frac{r_+ R}{l^2}}}\left[-\frac{\left( \cosh{\frac{r_+ R}{l^2}} - \cosh{\hat{t}}\cosh{\hat{\phi}}\right)}{l}\partial_{{\phi}} + \sinh{\hat{t}}\,\sinh{\hat{\phi}} \,\partial_{{t}}\right]\,.
\end{equation}
Clearly for a constant time slice $t=0$ at the boundary, (\ref{@107}) gives $(\cosh{\frac{r_+ R}{l^2}} - \cosh{\hat{\phi}})=0$, which implies that  $\phi =\pm R/l$ points remain invariant under the boundary modular flow generated by the boundary modular Hamiltonian.
\vspace{1mm}

\noindent
\emph{(ii) Timelike intervals:} Similarly, for a purely timelike subregion stretched between $P ( {\phi} = 0, {t} = - T_0 )$ and $Q ( {\phi} = 0, {t}= +T_0)$, the modular Hamiltonian and momentum take the forms
\begin{equation}
    \tilde{H}_{mod}^{(T)}:=\frac{\beta}{\sinh{\frac{r_+ T_0}{l^2}}} \bigg[(L_1 +L_{-1}  ) -  ( \bar{L}_1+\bar{L}_{-1}) - 2 \cosh{\frac{r_+ T_0 }{l^2}}(L_0 - \bar{L}_0)\bigg]\qquad\text{and}
\end{equation}
\begin{equation}
    \tilde{P}_{mod}^{(T)}:=\frac{\beta}{\sinh{\frac{r_+ T_0}{l^2}}} \bigg[(L_1 + L_{-1} ) +  ( \bar{L}_1 +\bar{L}_{-1}) - 2 \cosh{\frac{r_+ T_0}{l^2}}(L_0 + \bar{L}_0)\bigg]
\end{equation}
Once again, using \eqref{@m58} and \eqref{@m59}, these take the form
\begin{equation}
    \tilde{H}_{mod}^{(T)} := \frac{\beta}{\sinh{\frac{r_+ T_0}{l^2}}}\left[\frac{\left( \cosh{\frac{r_+ T_0}{l^2}} - \cosh{\hat{t}}\,\cosh{\hat{\phi}}\right)}{l}\partial_{{\phi}} - \sinh{\hat{t}}\,\sinh{\hat{\phi}}\, \partial_{{t}}\right]\qquad\text{and}\label{@113}
\end{equation}    
\begin{equation}
    \tilde{P}_{mod}^{(T)} := \frac{\beta}{\sinh{\frac{r_+ T_0}{l^2}}}\left[-\left( \cosh{\frac{r_+ T_0}{l^2}} - \cosh{\hat{t}}\,\cosh{\hat{\phi}}\right)\partial_{{t}} + \frac{1}{l}\sinh{\hat{t}}\,\sinh{\hat{\phi}}\, \partial_{{\phi}}\right]\,.
\end{equation}
Clearly the $\pm T_0$ endpoints are also invariant for the $\hat{\phi}=0$ slice. 

\subsubsection{Bulk flow: outside BTZ horizon}\label{appsec:modflowbtzout}

For the bulk points outside of horizon, the bulk isometry generators are respectively given by \cite{Keski-Vakkuri:1998gmz,Banerjee:2022jnv}
\begin{equation}
    L_{b,0} = - \partial_+\,, \hspace{0.5cm} \bar{L}_{b,0} = - \partial_-\,, \label{@m78}
\end{equation}
\begin{equation}
    L_{b,\pm1} = -\frac{\sqrt{r^2 - r_+^2}}{2r}\, e^{\pm\hat{x}^+}\Bigg(\frac{2r^2-r_+^2}{r^2-r_+^2}\partial_++\frac{r_+^2}{r^2-r_+^2}\partial_- \mp \frac{r}{l} \,\partial_r \Bigg)\,,\label{@m79}
\end{equation}
and
\begin{equation}
     \bar{L}_{b,\pm1} = -\frac{\sqrt{r^2 - r_+^2}}{2r}\, e^{\pm\hat{x}^-}\Bigg(\frac{2r^2-r_+^2}{r^2-r_+^2}\partial_-+\frac{r_+^2}{r^2-r_+^2}\partial_+ \mp \frac{r}{l}\, \partial_r \Bigg)\,.\label{@m80}
\end{equation} 
They are related to the bulk modular Hamiltonian and momentum in a manner similar to \eqref{@m9} and \eqref{@m11}.
\vspace{1mm}

\noindent
\emph{(i) Spacelike intervals:} For a spatial subregion at $t=0$ slice of size $2R$ (symmetric around $x=0$), we will now obtain
\begin{equation}
    \tilde{H}_{mod}^{bulk} := \frac{\beta}{\sinh{\frac{r_+ R}{l^2}}} \bigg[(L_{b,1} +L_{b,-1} ) +  ( \bar{L}_{b,1} +\bar{L}_{b,-1}) - 2 \cosh{\frac{r_+ R }{l^2}}(L_{b,0} + \bar{L}_{b,0})\bigg]\qquad\text{and}
\end{equation}
\begin{equation}
    \tilde{P}_{mod}^{bulk} := \frac{\beta}{\sinh{\frac{r_+ R}{l^2}}} \bigg[(L_{b,1} +L_{b,-1} ) -  ( \bar{L}_{b,1} +\bar{L}_{b,-1}) - 2 \cosh{\frac{r_+ R }{l^2}}(L_{b,0} - \bar{L}_{b,0})\bigg]\,.
\end{equation}
In particular, using \eqref{@m78}-\eqref{@m80}, the extended bulk modular Hamiltonian takes the form
\begin{eqnarray}
    \tilde{H}_{mod}^{bulk} := \frac{\beta}{\sinh{\frac{r_+ R}{l^2}}}\Bigg[\left( \cosh{\frac{r_+ R}{l^2}} - \frac{r}{\sqrt{r^2 - r_+^2}} \cosh{\hat{t}}\cosh{\hat{\phi}}\right)\partial_{{t}} - \frac{\sqrt{r^2-r_+^2}}{r\,l} \sinh{\hat{t}}\,\sinh{\hat{\phi}}\,\partial_{{\phi}}\nonumber\\
     + \frac{\sqrt{r^2 - r_+^2}}{l} \sinh{\hat{t}}\,\cosh{\hat{\phi}}\,\partial_r\Bigg]\,.\nonumber\\
     \label{@124}
\end{eqnarray}
We can see clearly that the surface defined by $\hat{t}=0$ and 
\begin{equation}
\sqrt{1-\frac{r_+^2}{r^2}} = \frac{\cosh{\frac{r_+ \phi}{l}}}{\cosh{\frac{r_+R}{l^2}}}
\end{equation}
will be invariant under the modular flow. This is the well-known equation for the RT surface for BTZ \cite{Kabat:2017mun}.

\vspace{1mm}

\noindent
\emph{(ii) Timelike intervals:} Similarly, for a timelike interval located symmetrically around $\hat{\phi}=0$ and of size $2T_0$, the resulting bulk operators are
\begin{equation}
    \tilde{H}_{mod}^{bulk(T)} := \frac{\beta}{\sinh{\frac{r_+ T_0}{l^2}}} \bigg[(L_{b,1} +L_{b,-1} ) -  ( \bar{L}_{b,1} +\bar{L}_{b,-1}) - 2 \cosh{\frac{r_+ T_0 }{l^2}}(L_{b,0} - \bar{L}_{b,0})\bigg]\qquad\text{and}
\end{equation}
\begin{equation}
    \tilde{P}_{mod}^{bulk(T)} := \frac{\beta}{\sinh{\frac{r_+ T_0}{l^2}}} \bigg[(L_{b,1} +L_{b,-1} ) +  ( \bar{L}_{b,1} +\bar{L}_{b,-1}) - 2 \cosh{\frac{r_+ T_0 }{l^2}}(L_{b,0} + \bar{L}_{b,0})\bigg]\,.
\end{equation}
Once again, using \eqref{@m78}-\eqref{@m80}, we find the modular Hamiltonian taking the following form
\begin{eqnarray}
	 \tilde{H}_{mod}^{bulk(T)}:= \frac{\beta}{\sinh{\frac{r_+ T_0}{l^2}}}\Bigg[\frac{\left( \cosh{\frac{r_+ T_0}{l^2}} - \frac{\sqrt{r^2 - r_+^2}}{r} \cosh{\hat{t}}\cosh{\hat{\phi}}\right)}{l}\partial_{{\phi}} - \frac{r}{\sqrt{r^2-r_+^2}} \sinh{\hat{t}}\sinh{\hat{\phi}}\,\partial_{{t}}\nonumber\\
	  + \frac{\sqrt{r^2 - r_+^2}}{l} \cosh{\hat{t}}\sinh{\hat{\phi}}\,\partial_r\Bigg]\,.\nonumber\\
	  \label{@127}
\end{eqnarray}
Clearly, for $\hat{\phi}=0$, the surface 
\begin{equation}\label{eq:slparttlbtz}
\sqrt{1-\frac{r_+^2}{r^2}} = \frac{\cosh{\frac{r_+T_0}{l^2}}}{\cosh{\frac{r_+ t}{l^2}}}
\end{equation}
will be invariant under the modular flow. This turns out to be the spacelike (green) part of the extremal surface as depicted in figure \ref{fig:HEE_02} (for points outside the horizon) \cite{Doi:2022iyj}.

\subsubsection{Bulk flow: inside BTZ horizon}\label{appsec:modflowbtzin}

For points inside the horizons of BTZ black hole we have the following bulk generators
\begin{equation}
    L_{b,0} =  \partial_+ \,, \hspace{0.5cm} \bar{L}_{b,0} = \partial_- \label{@m96}\,,
\end{equation}
\begin{equation}
   L_{b,\pm1} =\pm \frac{\sqrt{r_+^2 - r^2}}{2r} e^{\pm\hat{x}^+}\Bigg(\frac{2r^2-r_+^2}{r^2-r_+^2}\partial_++\frac{r_+^2}{r^2-r_+^2}\partial_- \mp \frac{r}{l} \partial_r \Bigg)\,,\label{@m97}
\end{equation}
and
\begin{equation}
    \bar{L}_{b,\pm1} = \pm\frac{\sqrt{r_+^2 - r^2}}{2r} e^{\pm\hat{x}^-}\Bigg(\frac{2r^2-r_+^2}{r^2-r_+^2}\partial_-+\frac{r_+^2}{r^2-r_+^2}\partial_+ \mp \frac{r}{l}  \partial_r \Bigg)\,.\label{@m98}
\end{equation}
These, accompanied by equations like \eqref{@m9} and \eqref{@m11} reproduce the corresponding actions due to modular Hamiltonians at the inside points.
\vspace{1mm}

\noindent
\emph{(i) Timelike intervals:} For our purposes it is sufficient to look into the modular flow inside BTZ due to a timelike subregion with endpoints  $({t} = -T_0, {t} = T_0)$ (for $\hat{\phi}=0$). In this case, we have
\begin{equation}
    \tilde{H}_{mod}^{bulk(T)} := \frac{\beta}{\sinh{\frac{r_+ T_0}{l^2}}} \bigg[(L_{b,1} +L_{b,-1} ) -  ( \bar{L}_{b,1} +\bar{L}_{b,-1}) - 2 \cosh{\frac{r_+ T_0 }{l^2}}(L_{b,0} - \bar{L}_{b,0})\bigg]\,,
\end{equation}
\begin{equation}
    \tilde{P}_{mod}^{bulk(T)} := \frac{\beta}{\sinh{\frac{r_+ T_0}{l^2}}} \bigg[(L_{b,1} +L_{b,-1} ) +  ( \bar{L}_{b,1} +\bar{L}_{b,-1}) - 2 \cosh{\frac{r_+ T_0 }{l^2}}(L_{b,0} + \bar{L}_{b,0})\bigg]\,.
\end{equation}
Using \eqref{@m96} to \eqref{@m98}, the extended timelike modular Hamiltonian takes the following form
\begin{eqnarray}
	    \tilde{H}_{mod}^{bulk(T)}:=\frac{\beta}{\sinh{\frac{r_+ T_0}{l^2}}}\Bigg[\frac{\left( - \cosh{\frac{r_+ T_0}{l^2}} - \frac{\sqrt{r_+^2 - r^2}}{r} \sinh{\hat{t}}\cosh{\hat{\phi}}\right)}{l}\partial_{{\phi}} - \frac{r}{\sqrt{r_+^2-r^2}} \cosh{\hat{t}}\sinh{\hat{\phi}}\,\partial_{{t}}\nonumber\\ +\frac{\sqrt{r_+^2 - r^2}}{l} \sinh{\hat{t}}\sinh{\hat{\phi}}\,\partial_r\Bigg]\,.\nonumber\\
	    \label{@137}
\end{eqnarray}
In this case as well, we see that for $\hat{\phi}=0$ slice, 
\begin{equation}\label{eq:btzext}
	\sqrt{\frac{r_+^2}{r^2}-1} = -\frac{\cosh{\frac{r_+T_0}{l^2}}}{\sinh{\frac{r_+ t}{l^2}}}
\end{equation}
surface is invariant under the resulting bulk modular flow. This is again the spacelike part of the extremal surface (green lines in figure \ref{fig:HEE_02}, but inside the horizon)  associated with the timelike boundary interval. Hence, once again, only the spacelike parts of the surface remain invariant under the bulk modular flow, but not the timelike part (red line in figure \ref{fig:HEE_02}). This will play a key role in the bulk reconstruction for BTZ background.

\section{Further details for outside BTZ bulk reconstruction}\label{app:outsidebtztechnical}

In this appendix, we will briefly sketch how to perform bulk reconstruction outside the BTZ black hole using the boundary constraints \eqref{eq:bdrycondoutbtz}. The entire analysis closely follows \cite{Kabat:2017mun}, with almost an interchange between spatial and temporal coordinates. 

To begin with, we need to define a slightly modified $p,q$ variables as
\begin{equation}
    q=t-\frac{il^2y}{r_+}+\frac{l^2 x}{r_+} \;,\; p=t+\frac{il^2y}{r_+}+\frac{l^2 x}{r_+}\,,
\end{equation}
and start with an ansatz for the bulk scalar field at a point $Q$ as 
\begin{equation}\label{eq:outbtzphiansatz}
    \Phi(Q)=\int d\hat{q}\,d\hat{p}\;g(\hat{q},\hat{p})\,\mathcal{O}(\hat{q},\hat{p})\,.
\end{equation}
where for a variable $x$, $\hat{x}=\frac{r_+ x}{l^2}$.
Starting with
\begin{equation}
    [\tilde{H}_{B}^{(T)},\Phi(Q)]=0
\end{equation}
and doing integration by parts, we get an equation
\begin{align}\label{247}
\bigl[(\cosh\hat{T}_B-\cosh(\hat{p}-\hat{t}_0))\,\partial_{\hat{p}}&-(\cosh\hat{T}_B-\cosh(\hat{q}-\hat{t}_0))\,\partial_{\hat{q}}\nonumber\\
&+(h-1)(\sinh(\hat{p}-\hat{t}_0)-\sinh(\hat{q}-\hat{t}_0)) \bigr] g(\hat{q},\hat{p})=0\,.
\end{align}
To obtain $[\tilde{H}_{A}^{(T)},\Phi(Q)]$, we simply need to put $t_0=0$ and replace $T_B$ by $T_A$ (the half-length of the time interval $A$). Solving $[\tilde{H}_{A}^{(T)},\Phi(Q)]=0$ using method of characteristics, we obtain a general solution ($a_0$ below is a constant)
\begin{equation}
       g(\hat{q},\hat{p})=a_0 f(x)U^{h-1}\,,
\end{equation}
where 
\begin{equation}
       x=\frac{\sinh\left(\frac{\hat{T}_A+\hat{q}}{2}\right)\sinh\left(\frac{\hat{T}_A+\hat{p}}{2}\right)}{\sinh\left(\frac{\hat{T}_A-\hat{q}}{2}\right)\sinh\left(\frac{\hat{T}_A-\hat{p}}{2}\right)}
\end{equation}
and 
\begin{equation}
       U=\sinh\left(\frac{\hat{T}_A+\hat{q}}{2}\right)\,
        \sinh\left(\frac{\hat{T}_A+\hat{p}}{2}\right)\,
        \sinh\left(\frac{\hat{T}_A-\hat{q}}{2}\right)\,
        \sinh\left(\frac{\hat{T}_A-\hat{p}}{2}\right)\,.
\label{248}
\end{equation}
One can now impose \([\tilde{H}_{B}^{(T)},\Phi(Q)]=0\) with the solution obtained above in order to find out the function $f(x)$. After some algebra, one can see that $f(x)$ satisfies the following equation
\begin{equation}
    \frac{df}{dx}=\frac{h-1}{x}\frac{x-\alpha}{x+\alpha}\,f\,,
\end{equation}
with
\begin{equation}
\alpha=\frac{\sinh{\hat{t}_0}\sinh\hat{T}_A+\cosh{\hat{t}_0}\cosh\hat{T}_A-\cosh\hat{T}_B}{\sinh{\hat{t}_0}\sinh\hat{T}_A-\cosh{\hat{t}_0}\cosh\hat{T}_A+\cosh\hat{T}_B}\,.
\end{equation}
The above equation has a solution
\begin{equation}
   f(x)=c_1\left(\frac{(x+\alpha)^2}{x}\right)^{h-1}\,.
\end{equation}
In order to obtain the solution in the required form, one can define a new variable $\hat{t}_*$, which is nothing but the time at point $Q$. This definition has been written using hindsight; in other words, it can be obtained by solving for the intersection point of two spacelike parts of the extremal surfaces in the BTZ background \eqref{eq:slparttlbtz}. One part stretching from the top point of interval $A$, and the other from the bottom point of interval $B$ thereby giving rise an emergent bulk direction (see figure \ref{fig:BTZ_02}). In terms of this new variable, we have
\begin{equation}
       \tanh\hat{t}_*=\frac{1}{\sinh\hat{t}_0}\left(\cosh\hat{t}_0-\frac{\cosh\hat{T}_B}{\cosh\hat{T}_A}\right)\,.
\end{equation}
In terms of this $\hat{t}_*$, $\alpha$ and $g(\hat{q},\hat{p})$ take the following values
\begin{equation}
   \alpha=\frac{\cosh\hat{t}_*\sinh\hat{T}_A+\sinh\hat{t}_*\cosh\hat{T}_A}{\cosh\hat{t}_*\sinh\hat{T}_A-\sinh\hat{t}_*\cosh\hat{T}_A}\qquad\text{and}
\end{equation} 
\begin{align}
       g(\hat{q},\hat{p})&=c_1\,\left(\cosh\left(\frac{\hat{p}+\hat{q}}{2}-\hat{t}_*\right)-\frac{\cosh\hat{T}_A}{\cosh\hat{t}_*}\cosh\left(\frac{\hat{p}-\hat{q}}{2}\right)\right)^{\Delta-2}\nonumber\\
       &=c_1\,\left(\cos{y}-\frac{\cosh\hat{T}_A}{\cosh\hat{t}_*}\cosh{x} \right)^{\Delta-2}\,. \label{x}
\end{align} 
\eqref{x} is indeed the smearing function of a local bulk field \( \Phi(r,\phi=0,t_*)\) outside the horizon of a BTZ black hole if the bulk field is located at a radial value $r$ given by \eqref{eq:slparttlbtz}. In other words, we have an emergent bulk direction arising out of boundary constraints. The fact that the boundary support in \eqref{eq:outbtzphiansatz} is over a spacelike separation $\cos{y}>\frac{\cosh\hat{T}_A}{\cosh\hat{t}_*}\cosh{x}$, arises from requiring that there are no boundary terms when we integrate by parts in \eqref{247}. 

\section{Further details on interior BTZ reconstruction}\label{app:insidebtztechnical}

To calculate \([\tilde{H}_R^{(T)},\Phi_L]\), we will need to realize the thermofield double (TFD) structure of the CFT state \eqref{eq:TFD} (or, in the bulk, that of the BTZ black hole) \cite{Fidkowski:2003nf} which implies that at the level of expectation values
\begin{equation}\label{eq:kms}
	\langle\mathcal{O}^L(\hat{t},\hat{\phi})\dots\rangle_{TFD}\equiv \langle\mathcal{O}^R(-\hat{t}-\frac{i\hat{\beta}}{2},\hat{\phi})\dots\rangle_{TFD}\,.
\end{equation}
Of course, from boundary perspective $\hat{\beta}$ is just $2\pi$, which will ultimately provide the HKLL BTZ fields for a certain choice of $\hat{\beta}$, namely for $\hat{\beta}=\frac{r_+\beta}{l^2}$, where $\beta$ is the inverse temperature of the BTZ black hole. This is sketched in figure \ref{fig:BTZ_04} below. This essentially takes the operators from right boundary to the left and vice-versa \cite{Haag:1992hx,Borchers:2000pv,Summers:2003tf,Witten:2018lha}.
 \begin{figure}[h]
    \centering
    \includegraphics[scale=0.5]{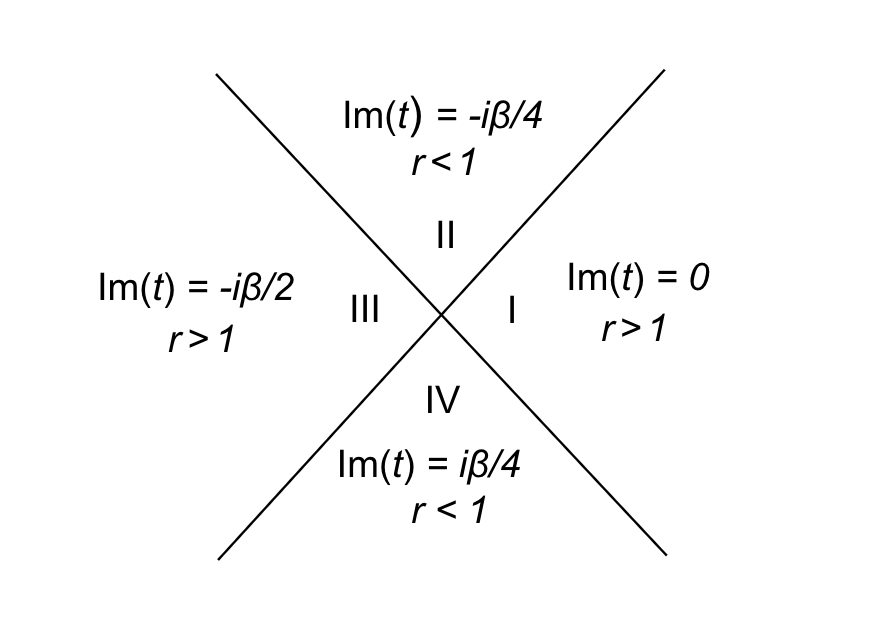}
    \caption{Analytic continuations (in time) between different quadrants of BTZ black hole.}
    \label{fig:BTZ_04}
\end{figure}
Remembering the HKLL formula for the interior fields
\begin{align}
   & \Phi(\phi,t,r)\nonumber\\
   &= c_{\Delta}\left[ \int_{\sigma>0} dy\,dx\, \left(\frac{\sigma}{r'}\right)^{\Delta-2} \,\mathcal{O}^{R}\left(\phi + \frac{ily}{r_+},t+\frac{l^2 x}{r_+}\right)+\int_{\sigma<0}dy\,dx\, \left(\frac{-\sigma}{r'}\right)^{\Delta-2} (-1)^{\Delta}\mathcal{O}^{L}\left(\phi + \frac{ily}{r_+},t+\frac{l^2 x}{r_+}\right)\right]\nonumber\\
   &=\Phi_R+\Phi_L\,, \label{27inapp}
\end{align}
and that the relevant timelike modular Hamiltonian is 
\begin{equation}
    \tilde{H}_R^{(T)}=\tilde{c} \int_{-\infty}^{\infty}d\xi\, \left(\cosh\left(\frac{r_+ T_0}{l^2}\right)-\cosh\left(\frac{r_+ \xi}{l^2}\right) \right)T_{\xi{\xi}} + \text{anti-chiral part}\,, \label{26app}
\end{equation}
we have already written down $\left[\tilde{H}_{R}^{(T)},\Phi_R(\phi=0,t,r)\right]$ in \eqref{33in}. For the second term, we can use \eqref{eq:kms} above and define new lightcone variables related to the usual ones (in \eqref{30}) in the following way\footnote{In writing the formulas below and applying them in HKLL equations, we have in mind evaluating correlators as in \eqref{eq:kms}.} 
\begin{equation}
    \xi^{new} = -t-\frac{il^2y}{r_+}-\frac{i\pi l^2}{r_+}=-\bar{\xi}-\frac{i\pi l^2}{r_+} \label{38}
\end{equation} 
and 
\begin{equation}
    \bar{\xi}^{new}=-t+\frac{il^2y}{r_+}-\frac{i\pi l^2}{r_+}=-\xi-\frac{i\pi l^2}{r_+}\,. \label{39}
\end{equation} 
Using these new set of variables, we find that the $\left[\tilde{H}_{R}^{(T)},\mathcal{O}_L(\xi,\bar{\xi})\right]$ can be evaluated to be
\begin{align}
	&[\tilde{H}_R^{(T)},\mathcal{O}_{R}(\xi^{new},\bar{\xi}^{new})]\nonumber\\
	&=\tilde{c}\, \Biggl(-\frac{\Delta r_+}{2l^2}\left(\sinh\left(\frac{r_+\bar{\xi}^{new}}{l^2}\right)-\sinh\left(\frac{r_+\xi^{new}}{l^2}\right)\right)-\biggl(\cosh\left(\frac{r_+T_0}{l^2}\right)-\cosh\left(\frac{r_+\bar{\xi}^{new}}{l^2}\right)\biggr)\partial_{\bar{\xi}^{new}}\nonumber\\
	& +\biggl(\cosh\left(\frac{r_+T_0}{l^2}\right)-\cosh\left(\frac{r_+\xi^{new}}{l^2}\right)\biggr)\partial_{\xi^{new}}\biggr) \mathcal{O}(\xi^{new},\bar{\xi}^{new}) \label{40}
\end{align}
Finally, redefining $(q,p)$ slightly (compared to \eqref{30}) by
\begin{equation}
    p^{new}=\xi^{new}+\frac{l^2x}{r_+} \qquad\text{and}\qquad q^{new}=\bar{\xi}^{new}+\frac{l^2x}{r_+}\,, \label{42}
\end{equation}
we have
\begin{align}
	&[\tilde{H}_R^{(T)},\mathcal{O}_{R}(\xi^{new},\bar{\xi}^{new})]\nonumber\\
	&=\tilde{c}\,\Biggr(\frac{\Delta r_+}{2l^2}\biggl(\sinh\left(\frac{r_+q^{new}}{l^2}\right)-\sinh\left(\frac{r_+p^{new}}{l^2}\right)\biggr)-\biggl(\cosh\left(\frac{r_+T_0}{l^2}\right)-\cosh\left(\frac{r_+q^{new}}{l^2}\right)\biggr)\partial_{q^{new}}\nonumber\\
	& +\biggl(\cosh\left(\frac{r_+T_0}{l^2}\right)-\cosh\left(\frac{r_+p^{new}}{l^2}\right)\biggl)\partial_{p^{new}}\biggr) \mathcal{O}(q^{new},p^{new})\,. \label{43}
\end{align}
Putting everything together, we have
\begin{align}
	&\left[\tilde{H}_{R}^{(T)},\Phi_L(\phi=0,t,r)\right]\nonumber\\
	&=C\int_{\sigma<0} dq\,dp\, \Biggl(\cosh \left(\frac{r_+(q^{new}-p^{new})}{2l^2} \right)+\sqrt{\frac{r_+^2}{r^2}-1}\, \sinh \left(\frac{r_+(p^{new}+q^{new}+2t+
	\frac{2i l^2\pi}{r_+})}{2l^2}\right) \Biggr)^{\Delta-2}\nonumber\\
	& \Bigg(\frac{\Delta r_+}{2l^2}\left(\sinh\left(\frac{r_+q^{new}}{l^2}\right)-\sinh\left(\frac{r_+p^{new}}{l^2}\right)\right)-\left(\cosh\left(\frac{r_+T}{l^2}\right)-\cosh\left(\frac{r_+q^{new}}{l^2}\right)\right)\partial_{q^{new}}\nonumber\\
	&+\left(\cosh\left(\frac{r_+T_0}{l^2}\right)-\cosh\left(\frac{r_+p^{new}}{l^2}\right)\right)\partial_{p^{new}}\Bigg) \mathcal{O}^{R}(q^{new},p^{new}) \label{44}
\end{align}
with $C=c_\Delta\, \tilde{c}\,(-1)^{2\Delta-2}$. This time
\begin{equation}\label{eq:intcommsapp}
[\tilde{H}_R^{(T)},\Phi_R]+ [\tilde{H}_R^{(T)},\Phi_L]=0
\end{equation}
gives (adding the above to \eqref{33in})
\begin{align}
	&\left(\sqrt{\frac{r_+^2}{r^2}-1}\sinh\left(\frac{r_+t}{l^2}\right)+\cosh\left({\frac{r_+T_0}{l^2}}\right)\right) \Biggr(\int_{\sigma>0} \left(\frac{\sigma_-}{r'}\right)^{\Delta-3}\sinh\left(\frac{r_+(p-q)}{l^2}\right)\,\mathcal{O}^R(q,p)\nonumber\\
	&-\int_{\sigma<0}(-1)^{2\Delta-2}\left(\frac{\sigma_+}{r'}\right)^{\Delta-3}\sinh\left(\frac{r_+(q^{new}-p^{new})}{l^2}\right)\mathcal{O}^R(q^{new},p^{new})\Biggr)=0\,.
\end{align}
Once again this generically implies the same geodesic equation
\begin{equation}
    \sqrt{\frac{r_+^2}{r^2}-1}=-\frac{\cosh({\frac{r_+T_0}{l^2}})}{\sinh(\frac{r_+t}{l^2})}\,. \label{45app}
\end{equation}

\bibliographystyle{utphys}
\bibliography{Hmod}
\end{document}